\documentclass{article}

\usepackage{mathptmx,helvet,courier,graphicx,makeidx,multicol} 
\usepackage[bottom]{footmisc}
\usepackage{color}\usepackage{float}

\renewcommand{\vec}[1]{\mathbf{#1}}
 
\usepackage{enumitem}
\usepackage{amssymb}
\usepackage{hyperref}

\usepackage{amsthm}
\newtheorem{remark}{Remark}

\begin{document}

%
\title{Continuation for thin film hydrodynamics and related scalar
  problems}
\author{S. Engelnkemper, S. V. Gurevich, H. Uecker, D. Wetzel, U. Thiele}

\maketitle

\abstract{This chapter illustrates how to apply continuation
  techniques in the analysis of a particular class of nonlinear
  kinetic equations that describe the time evolution of a single scalar field like a density or interface
  profiles of various types. We first systematically introduce these
  equations as gradient dynamics combining mass-conserving and
  nonmass-conserving fluxes followed by a discussion of nonvariational
  amendmends and a brief introduction to their analysis by numerical
  continuation.\\
  The approach is first applied to a number of common examples of
  variational equations, namely, Allen-Cahn- and Cahn-Hilliard-type
  equations including certain thin-film equations for partially
  wetting liquids on homogeneous and heterogeneous substrates as well
  as Swift-Hohenberg and Phase-Field-Crystal equations. Second we
  consider nonvariational examples as the Kuramoto-Sivashinsky
  equation, convective Allen-Cahn and Cahn-Hilliard equations and
  thin-film equations describing stationary sliding drops and a
  transversal front instability in a
  dip-coating.\\
  Through the different examples we illustrate how to employ the
  numerical tools provided by the packages \textsc{auto07p} and
  \textsc{pde2path} to determine steady, stationary and time-periodic
  solutions in one and two dimensions and the resulting bifurcation
  diagrams. The incorporation of boundary conditions and integral side
  conditions is also discussed as well as problem-specific
  implementation issues.\\[1em]
  \emph{Published as:} Engelnkemper, S., Gurevich, S.V., Uecker, H., Wetzel, D., Thiele, U.: Continuation for thin film hydrodynamics and related scalar problems. In: Gelfgat, A. (ed.): Computational Modelling of Bifurcations and Instabilities in Fluid Dynamics, Springer (2019).
}


\section{Introduction} \label{sec:intro}
The techniques of path-continuation are widely employed to obtain
branches of different types of solutions to nonlinear equations and,
in consequence, to numerically construct bifurcation diagrams
\cite{AllgowerGeorg1987,KrauskopfOsingaGalan-Vioque2007,Kuznetsov2010}.
A classical, quite versatile, often used package is
\textsc{auto07p}. It implements pseudo-arclength continuation
algorithms and uses a discretization in the time domain (for
time-periodic solutions and boundary value problems) based on
orthogonal collocation employing piecewise polynomials with a small
number of collocation points per interval of the adaptive mesh
\cite{DoedelOldeman2009}.  In this way, the package covers most
problems encountered for systems of ordinary differential equations
(ODE) \cite{DoKK1991ijbc,DoKK1991ijbcb} and has in some cases also been
used for selected partial differential equation (PDE) or
integro-differential equation (IDE) problems
\cite{BoEn2007pd,KoTh2014n,PAST2011pre}.

However, user-friendly tools to systematically apply continuation
techniques to PDE problems are still scarce. One recently developed
example is \textsc{pde2path}, like \textsc{auto07p} using
arclength continuation, but aiming at systems of PDEs in 1d, 2d and 3d,
with the spatial discretization based on finite elements (FEM). See also
\S\ref{sec:cont} for a brief review of arclength continuation and
further comments on available packages.

For spatially extended systems, e.g., described by evolution equations
for concentrations, densities or interface positions the issue of
symmetries and additional constraints like mass conservation becomes
important as they imply that additional side conditions have to be
imposed on the continuation path. This is well possible in both,
\textsc{auto07p} and \textsc{pde2path}, and normally results in
additional continuation parameters besides the main one.

Here, we illustrate how to apply continuation techniques in the
analysis of the solution behavior of a particular class of nonlinear kinetic
equations that describe the time evolution of a single scalar field
$\phi(\vec{r},t)$ by the action of mass-conserving fluxes
$\vec{j}_\mathrm{c}$ and nonmass-conserving fluxes (or rates) $j_\mathrm{nc}$, i.e.,
\begin{equation}
\partial_t \phi = -\nabla\cdot\vec{j}_\mathrm{c} + j_\mathrm{nc},
\label{eq:kinetic}
\end{equation}
where $\phi$ might, e.g., be a density,
concentration or film height, and ``mass'' is given by $m=\int \phi \, \mathrm{d}^n r$ for $n$ spatial dimensions. 
In many cases we first search steady solutions of (\ref{eq:kinetic}), 
fulfilling 
\begin{equation}
0=-\nabla\cdot\vec{j}_\mathrm{c} + j_\mathrm{nc}.
\label{kins}
\end{equation}
Note that all our equations are given in nondimensional form. However, sometimes we keep nondimensional parameters that could be eliminated by scaling in order to better keep track of the individual terms.
Transport equations such as (\ref{eq:kinetic}) often describe the time evolution of density and
interface profiles of various types. In many cases, the
fluxes can be written in gradient dynamics (or ``variational'') form. In the
treated case of a single scalar field $\phi$, such mass-conserving and
nonmass-conserving variational fluxes are given by
\begin{equation}
\vec{j}_\mathrm{c}^\mathrm{g} =  -Q
_\mathrm{c}\nabla\mu^\mathrm{g} = -Q_\mathrm{c}\nabla\frac{\delta \mathcal{F}[\phi]}{\delta\phi} 
\qquad\mathrm{and}\qquad 
j_\mathrm{nc}^\mathrm{g} = -Q_\mathrm{nc}\mu^\mathrm{g}=
-Q_\mathrm{nc}\frac{\delta \mathcal{F}[\phi]}{\delta\phi}, 
\label{eq:fluxes}
\end{equation}
respectively, i.e., Eq.~(\ref{eq:kinetic}) becomes
\begin{equation}
\partial_t \phi =
\nabla\cdot\left[Q_\mathrm{c}\nabla\frac{\delta \mathcal{F}[\phi]}{\delta\phi} \right] 
-Q_\mathrm{nc}\frac{\delta \mathcal{F}[\phi]}{\delta\phi}.
\label{eq:kinetic2}
\end{equation}
Here, $\mathcal{F}[\phi]$ is an appropriately defined energy functional (here we
request that it is bounded from below, i.e., exclude terms like
$x\phi^n$ - but cf.~Ref.~\cite{EWGT2016prf}) and the
variational derivative $\mu^\mathrm{g}=\delta \mathcal{F}/\delta\phi$ normally corresponds to a
chemical potential or pressure. The $Q_i$ are positive definite mobility
functions. Without additional nongradient terms, $\mathcal{F}$ corresponds to a Lyapunov
functional for the dynamics as $\mathrm{d}\mathcal{F}/\mathrm{d}t\le0$ \cite{thie2010jpcm}. 

The simplest such equation where a spatial structure evolves is the
diffusion equation -- a conserved gradient dynamics where $\mathcal{F}$ is the purely entropic free energy and
$Q_\mathrm{c}\sim\phi$ \cite{Doi2013}. As it is linear and only has a homogeneous steady
state, we do not consider it here. A related equation with nontrivial
steady states is the Allen-Cahn (AC) equation
that represents a nonconserved gradient dynamics
[Eq.~(\ref{eq:kinetic2}) with $Q_\mathrm{c}=0$] with
\begin{equation}
 \mathcal{F}[\phi]\,=\,\int\left[\frac{\sigma}{2}(\nabla \phi)^2 + f(\phi)\right]\,\mathrm{d}V
\label{eq:energyAC}
\end{equation}
where $f(\phi)$ is a local energy function that only depends on the
field itself but not on its derivatives. It is often a cubic or
quartic polynomial. Furthermore, $\sigma>0$ is a stabilizing interface stiffness, e.g., penalizes strong concentration gradients.

The AC equation can describe ordering dynamics
related to a structural phase transition (e.g., the coarsening
kinetics of crystal grains in alloys \cite{AlCa1979amm}), or simple
one-component reaction-diffusion systems. Depending on the type of
polynomial and scientific lineage, other names for this equation are
Nagumo equation, Fisher equation, Kolmogorov-Petrovsky-Piskunov (KPP)
equation, Fisher-Kolmogorov equation or Fisher-KPP equation
\cite{DoKK1991ijbcb,Saar1988pra}. Note that for complex $\phi$ it is 
equivalent to the Ginzburg-€"Landau equation.

Employing the energy functional (\ref{eq:energyAC}) in a conserved
gradient dynamics [Eq.~(\ref{eq:kinetic2}) with $Q_\mathrm{nc}=0$]
with $f$ being a quartic polynomial gives the Cahn-Hilliard (CH)
equation that describes e.g. the decomposition dynamics of a binary
mixture \cite{CaHi1958jcp,Cahn1965jcp}. Steady states of AC and CH
equations are related as discussed more in detail in
section~\ref{sec:ac-ch}. Note that the thin-film (TF) equation
describing the dewetting of a liquid thin film or droplet coarsening
on a horizontal solid substrate
\cite{ordb1997rmp,thie2007,BEIM2009rmp,thie2010jpcm} also corresponds
to such a conserved gradient dynamics with a nonlinear (cubic)
$Q_\mathrm{c}$ and particular functions $f(\phi)$ representing various
wetting potentials
\cite{Mitl1993jcis,ShJa1994jcsft,PiPo2000pre,ThVN2001prl}.

The square-gradient term in the energy~(\ref{eq:energyAC}) penalizes
interfaces and acts stabilizing in AC and CH dynamics. It can also have a
negative prefactor and then promotes interface formation. In this case, normally, a higher
order term acts stabilizing. The simplest corresponding energy functional has the form
\begin{equation}
\mathcal{F}[\phi]\,=\,\int\left[\frac{\kappa}{2}(\Delta \phi)^2 - \frac{\sigma}{2}(\nabla \phi)^2
+ f(\phi) \right]\mathrm{d}V
\label{eq:en2}
\end{equation}
where, in addition to eq.~(\ref{eq:energyAC}), $\kappa>0$ represents the energetic cost of corners
in the profile, and $f(\phi)$ is a local free energy as above.

A nonconserved gradient dynamics on the energy functional
(\ref{eq:en2}) is the Swift-Hohenberg (SH) equation that is the
typical model equation for the dynamics of pattern formation
close to the threshold of a short-scale instability
\cite{CrHo1993rmp}. It is usually applied with $f(\phi)$ being a
polynomial up to sixth order
\cite{SaBr1996pd,BuKn2006pre,ALBK2010sjads,MaSa2014pd}.
Employing the energy functional (\ref{eq:en2}) in a conserved gradient dynamics
gives the Phase-Field-Crystal (PFC) equation
\cite{ElGr2004pre,ELWG2012ap} (also called conserved Swift-Hohenberg
(cSH) equation \cite{TARG2013pre}) that describes e.g.\ the
microscopic crystallization dynamics of a colloidal suspension on
diffusive timescales \cite{ARTK2012pre} and is also extensively used
in material science \cite{ElGr2004pre,TTPT2010jpm,ELWG2012ap}.  Steady
states of SH and PFC equations are discussed in more detail in
section~\ref{sec:sh-pfc}.

Note that there exist many more kinetic equations of the form
(\ref{eq:kinetic}). Here, we only mention classical Dynamical Density
Functional Theories (DDFT) describing the behavior of colloidal
systems \cite{MaTa2000jpm,ArRa2004jpag,ELWG2012ap}. They normally
correspond to a conserved dynamics ($Q_\mathrm{nc}=0$) with a
functional $\mathcal{F}[\phi]$ that contains nonlocal kernels. Then
Eq.(\ref{eq:kinetic2}) corresponds to an integro-differential equation
(IDE).  

The study of nongradient variants of all the introduced equations is
widespread, but to our knowledge no overall systematics has been
worked out. In such equations both fluxes in Eq.~(\ref{eq:kinetic})
are sums of a gradient dynamics (or ``variational'') term as given in
(\ref{eq:fluxes}) and a nongradient dynamics (``nonvariational'')
term. In the context of Eq.~(\ref{eq:kinetic2}) the nonvariational contribution
typically takes three forms as summarized in the general
evolution equation
\begin{equation}
\partial_t \phi = \nabla \cdot\left[ Q _\mathrm{c}\nabla\left(\frac{\delta
    \mathcal{F}[\phi]}{\delta\phi} +\mu_\mathrm{c}^\mathrm{ng}\right)- \vec{j}_\mathrm{c}^\mathrm{ng} \right]
-\left( Q_\mathrm{nc}\frac{\delta \mathcal{F}[\phi]}{\delta\phi} +
  \mu_\mathrm{nc}^\mathrm{ng} \right).
\label{eq:kinetic:full}
\end{equation}
These nongradient contributions are related to
(i) an additional nonequilibrium chemical potential
$\mu_\mathrm{c}^\mathrm{ng}$ in the conserved dynamics, (ii) an additional
flux $\vec{j}_\mathrm{c}^\mathrm{ng}$ generated by an additional driving
force that can not be written as the gradient of a chemical potential, and
(iii) a nonequilibrium chemical potential $\mu_\mathrm{nc}^\mathrm{ng}$
in the nonconserved dynamics.

(i) The first type are additions $\mu_\mathrm{c}^\mathrm{ng}$ to the
chemical potential in the mass-conserving contribution that can not be
obtained by a variation of an energy functional. Up to second order
they are of the form
$(1-\tilde\beta)\frac{\nu'(\phi)}{2}|\nabla\phi|^2 +(1+\beta)
\nu(\phi)\Delta\phi$ where $\nu(\phi)$ is some function. It is
nonvariational for all $\beta,\tilde\beta\neq0$ (that may depend on
$\phi$). For the Cahn-Hilliard equation, common choices are
$\beta=\tilde\beta=-1$ \cite{WTSA2014nc,CaTa2015arcmp} and
$\beta=\tilde\beta =1$ \cite{STAM2013prl}.

(ii) The second type are additions $\vec{j}_\mathrm{c}^\mathrm{ng}$ to
the mass-conserving flux that break the isotropy (in 1d parity) of the
system and correspond to lateral driving forces. Typical contributions
are of the form $(\phi^n,0)^T$ with $n$ being some integer. For
instance, $n=1$ results in a driving comoving frame term relevant
e.g.\ in dragged-plate studies of Landau-Levich type (TF equation) or
descriptions of Langmuir-Blodgett transfer of surfactant molecules
onto a moving plate (convective CH equation) 
\cite{GTLT2014prl,KoTh2014n} or for driven pinned atomic monolayers
(PFC, \cite{ARKE2009pre}) or clusters (DDFT, \cite{PAST2011pre});
$n=2$ is used in another convective CH equation \cite{GNDZ2001prl},
in the Kuramoto-Sivashinsky (KS) equation \cite{HyNi1986pd}, (dissipative)
Burgers equation \cite{FrBe2001},
and liquid films driven by a thermal gradient \cite{Muen2003prl}; and
$n=3$ occurs in the thin-film equation for sliding drops or film flow
on an incline \cite{SLLS2007pf,EWGT2016prf}. Other terms are
dispersion contributions $\partial_{xx}\phi$ added to the flux in
one-dimensional KS equations \cite{AlDe1994pla}.

(iii) Also nonconserved nongradient terms
$\mu_\mathrm{nc}^\mathrm{ng}$ may be added to a mass-conserving
dynamics. For instance, $|\nabla\phi|^2$ is such a term in the KS
equation in the form used in \cite{KeNS1990sjam}\footnote{Note that
  nongradient terms $|\nabla\phi|^2$ and $\phi \nabla\phi$ are often
  related by a transformation that, however, also transforms conserved
  into nonconserved dynamics, therefore here we explicitly list both
  expressions.}. Such a term also appears in the Kardar-Parisi-Zhang
(KPZ) equation \cite{KaPZ1986prl} and is also used in a
nonvariational version of the SH equation \cite{HoKn2011pre}. One also finds
$\mu_\mathrm{nc}^\mathrm{ng}$-terms
$(1-\tilde\beta)\frac{\nu'(\phi)}{2}|\nabla\phi|^2 +(1+\beta)
\nu(\phi)\Delta\phi$ with $\nu=\phi$ and $\beta=\tilde\beta$ in AC \cite{ACGW2017pre} and SH
\cite{KoTl2007c,BuDa2012sjads} equations. Similar and higher order terms are used in Ref.~\cite{GoBo2007pre}. Another option are dispersion terms
proportional to $\partial_{xxx}\phi$ \cite{BuHK2009pre}.
A related subtle way to break the
gradient-dynamics structure is to combine conserved and nonconserved
fluxes that individually have gradient-dynamics form as in
(\ref{eq:fluxes}) but correspond to different energy functionals $\mathcal{F}$. However, the difference can
always be expressed as a particular form of
$\mu_\mathrm{nc}^\mathrm{ng}$. This occurs e.g.\
in some thin-film models with evaporation \cite{BeMe2006prl}. For
further discussion of evaporation terms see \cite{Thie2014acis}.

The chapter is structured as follows. In section~\ref{sec:cont} we
give a brief introduction to the general approach of numerical
continuation for kinetic equations as (\ref{eq:kinetic:full})
used in the framework of \textsc{pde2path} and \textsc{auto07p}. Section~\ref{sec:ac-ch}
focuses on steady states of the AC equation and the CH equation in 2d
as well as on steady states of a variational form of the TF equation for
partially wetting liquids on homogeneous and heterogeneous 2d
substrates, while section~\ref{sec:sh-pfc} discusses the SH equation
and a corresponding PFC model focusing on localized states in 1d and
2d. After the variational models selected nonvariational cases are
treated. Section~\ref{kssec} considers the steady and time-periodic
solutions of the KS equation in 1d, while \S\ref{sec:nonvar-ac} and
\S\ref{sec:nonvar-ch} study a convective AC equation and a
convective CH equation, respectively, in heterogeneous systems where
pinning effects compete with lateral driving (again steady and time-periodic
solutions in 1d). The  final result section ~\ref{sec:nonvar-tf}  focuses on  a general
nonvariational form of the TF equation studying stationary sliding
drops in 2d and a transversal front instability in a dip-coating (or
Landau-Levich) geometry.
We summarize and give an outlook in the concluding section~\ref{sec:summ}.


\section{Continuation approach} \label{sec:cont}
We briefly review the (pseudo-)arclength continuation method to find solution branches and bifurcations for nonlinear equations depending on parameters. One important advantage over natural continuation is the ability to follow solution branches through saddle-node bifurcations, essential for many of the example systems analyzed in this chapter. The approach follows \cite{keller77}, see also \cite{K86}.  
Much of the theory can 
be formulated in general Banach spaces, but for simplicity we restrict ourselves
to finite dimensions, and consider 
\begin{equation}
M\frac{\mathrm{d}}{\mathrm{d}t} u=-G(u,\lambda), \label{e0} 
\end{equation}
where $u=(u_1,\ldots,u_{n_u})\in\mathcal{X}=\mathbb{R}^{n_u}$ for instance 
denotes the nodal values of a spatial discretization 
of $\phi$ from (\ref{eq:kinetic}). 
In the following 
we assume for notational simplicity that the matrix $M\in\mathbb{R}^{n_u\times n_u}$ 
is the identity, but it may also be singular; 
in the context of FEM discretizations as in \textsc{pde2path}, it
corresponds to the 
so-called mass matrix. Note that $M$ 
plays no role for the computation of steady solutions of (\ref{e0}), 
but it influences the spectrum of the linearization $M\frac{\mathrm{d}}{\mathrm{d}t} v=-\partial_u G(u,\lambda)v$ 
around a steady solution. Further comments are given below.
In the simplest case $G\in C^1(\mathcal{X}\times\mathbb{R}, \mathcal{X})$ and 
$\lambda\in\mathbb{R}$ stands for a scalar ``active'' parameter, 
but often (\ref{e0}) must be 
extended by $n_q$ additional equations (constraints, or phase conditions), 
and to have one-dimensional continua of solutions we then need 
$n_q$ additional active parameters. 
We explain some general ideas combined with some comments on how they are 
applied in \textsc{auto07p} and \textsc{pde2path}. 

\subsection{Continuation of branches of steady solutions} 
\begin{figure}[htbp] \center
\includegraphics[width = 0.64\textwidth]{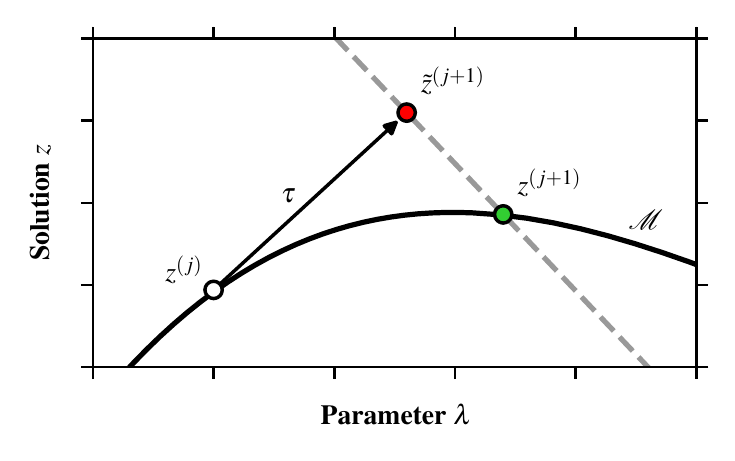}
\caption{Sketch of pseudo-arclength continuation in the plane spanned by
continuation parameter and solution measure.}
\label{fig:sketch}
\end{figure}

For later generality, i.e., for adding constraints, 
in our notations we assume 
that $\lambda\in\mathbb{R}^{n_p}$, $n_p\ge 1$, is a parameter vector, but 
first let $n_p=1$. 
We assume that $G\in C^1(\mathcal{X}\times\mathbb{R}, \mathcal{X})$ and consider an arc or ``branch''  $s\mapsto z(s):=(u(s),\lambda(s))\in \mathcal{X}\times\mathbb{R}$ of steady solutions 
of (\ref{e0}), parametrized by  $s\in\mathbb{R}$. We set up the extended system 
\begin{equation}
H(u,\lambda)=\left(\begin{array}{c} G(u,\lambda)\\p(u,\lambda,s)\end{array}\right)=0\in \mathcal{X}\times \mathbb{R}, \label{esys} 
\end{equation}
where $p$ is used to make $s$ an approximation to arclength on the 
solution arc. 
The standard choice is as follows.  Given 
 $s_0$ and a point $(u_0,\lambda_0):=(u(s_0),\lambda(s_0))$, and 
additionally a tangent vector 
$\tau_0:=(u_0',\lambda_0'):=\frac{{\rm d}}{{\rm d}s}(u(s),\lambda(s))|_{s=s_0}$ we use, 
for $s$ near $s_0$, 
\begin{equation}
p(u,\lambda,s):=\xi  \left\langle u_0',u(s)-u_0 \right\rangle  +(1-\xi)\left\langle\lambda_0',\lambda(s)-\lambda_0 \right\rangle  \label{pdef}
-(s-s_0). 
\end{equation}
Here $\left\langle u,v \right\rangle$ and $\left\langle\lambda,\mu \right\rangle$ are the inner products in $\mathbb{R}^{n_u}$ 
and $\mathbb{R}^{n_p}$, respectively, 
$0<\xi<1$ is a weight, and $\tau_0$ is assumed to be normalized in 
the weighted norm 
$$
\|\tau\|_\xi:=\sqrt{\left\langle\tau,\tau\right\rangle_\xi}, \qquad 
\left\langle\left(\begin{array}{c} u\\ \lambda\end{array} \right),\left(\begin{array}{c}  v\\ \mu\end{array}\right)\right\rangle_\xi:=\xi \left\langle u,v\right\rangle +(1-\xi)
\left\langle\lambda,\mu\right\rangle . 
$$
For fixed $s$, 
$p(u,\lambda,s)=0$ thus defines a hyperplane perpendicular 
(in the inner product $\left\langle\cdot,\cdot\right\rangle_\xi$) 
to $\tau_0$ at distance ${\rm ds}:=s-s_0$ from $(u_0,\lambda_0)$. 
A typical choice for $\xi$ is $\xi=n_p/n_u$ which gives the parameter (vector) $\lambda$ 
the same weight as the solution vector $u$ in the arclength. 
We then 
use a predictor $(u^1,\lambda^1)=(u_0,\lambda_0)+{\rm ds}\,\tau_0$ for a solution (\ref{esys}) 
on that hyperplane, followed by a corrector using Newton's method in the form 
\begin{equation}
\left(\begin{array}{c} u^{l+1}\\
\lambda^{l+1}\end{array}\right) =\left(\begin{array}{c} u^{l}\\\lambda^{l}\end{array}\right) - \mathcal{A}(u^l,\lambda^l)^{-1}H(u^l,\lambda^l), \quad 
\mathrm{ where\ } \mathcal{A}=\left(\begin{array}{cc} G_u & G_\lambda\\
\xi u_0' & (1-\xi)\lambda_0'\end{array}\right). \label{newton} 
\end{equation}
Here $\mathcal{A}^{-1}H$ stands for the solution of the linear system $\mathcal{A} z=H$ by 
a suitable method, i.e., $\mathcal{A}^{-1}$ should never be computed, except maybe for small $n_u$. 
Since $\partial_s p=-1$, on a smooth solution arc we have 
\begin{equation}
\displaystyle \mathcal{A}(s)\left(\begin{array}{c} u'(s)\\ \lambda'(s)\end{array}\right)=-\left(\begin{array}{c} 0\\\partial_s p\end{array}\right)=\left(\begin{array}{c} 0\\ 1\end{array}\right). 
\end{equation}
Thus, after convergence of (\ref{newton}) yields a new point $(u_1,\lambda_1)$ 
and Jacobian $\mathcal{A}$, the tangent direction $\tau_1$ at $(u_1,\lambda_1)$ 
with conserved orientation, i.e., sign$\left\langle\tau_0,\tau_1\right\rangle_\xi=1$, 
can be computed from 
$\displaystyle \mathcal{A}\tau_1=\left(\begin{array}{c} 0\\ 1\end{array}\right), \mathrm{with\ normalization } \|\tau_1\|_\xi=1. $
See Fig.~\ref{fig:sketch} for a sketch. 

The Jacobian $\mathcal{A}$ from (\ref{newton}) 
is nonsingular, except at steady bifurcation points (BPs), where 
two or more different branches of steady solutions intersect 
transversally. This follows from 
the so-called bordering lemma, see \cite[Chapter 3]{gov2000}. In particular, at so-called regular fold points, where the branch turns around, 
$G_u$ is singular but $\mathcal{A}$ is regular such that arclength continuation 
around fold points is no problem. Moreover, the predictor generically shoots beyond BPs, and the Newton method converges in cones around the branch 
with tips in the BPs, see, e.g., \cite[\S5.13]{K86}. 

\subsection{Bifurcations}
At BPs, eigenvalues $\mu(s)$ of $G_u$ must cross the imaginary 
axis. These can either be real eigenvalues, which under mild 
additional assumptions yields a steady bifurcation, 
or a complex conjugate pair $\mu_{\pm}(s_0)=\pm\mathrm{i} \omega(s_0)$ of eigenvalues, 
which under mild additional assumptions yields a Hopf bifurcation, 
i.e., the bifurcation of time periodic orbits ('Hopf orbits') 
for (\ref{e0}), with period near $2\pi/\omega(s_0)$ close to the bifurcation.   
To detect BPs, two simple methods which can also be used for large $n_u$, 
i.e., for (\ref{e0}) obtained from a spatial PDE discretization, are as 
follows. Here and in the following we always assume that we compare quantities 
evaluated at two points $z(s_1), z(s_2)$ on a branch, at distance $\mathrm{d}s$. 
Of course, all algorithms must fail if, e.g., a simple eigenvalue of 
$G_u(u(s))$ or $\mathcal{A}(s)$ 
moves back and forth between $s_1$ and $s_2$, e.g., if the step size $\mathrm{d}s$ is 
too large. 
\begin{itemize}
\item[(a)] Monitor sign changes of det$\mathcal{A}$. 
This is cheap if a lower upper ($\mathrm{LU}$) decomposition of $\mathcal{A}$ is available, which is 
usually the case if direct solvers  are used for the linear systems in 
(\ref{newton}). On the other hand, sign changes of det$\mathcal{A}$ 
only detect an odd number of eigenvalues crossing, and in particular 
cannot detect Hopf bifurcations. 
\item[(b)] Compute (by, e.g., inverse vector iteration) 
a (small) number of eigenvalues $\mu$ of $G_u$ closest to $0$ (and possibly also 
closest to spectral shifts $\mathrm{i} \omega_j$, $\omega_j\in\mathbb{R}$), and monitor 
the number of these eigenvalues with real parts greater than zero.  
This works well for dissipative problems where only few 
eigenvalues are close to the imaginary axis, and where, to detect 
Hopf bifurcations, guesses for suitable $\omega_j$ are available. A 
method to obtain such guesses, and ways in which the algorithm may 
fail, are explained in \cite[\S2.1]{hotheo}. 
\end{itemize}
After {\em detection} of a BP (or rather a possible BP, due to possibly 
spurious detection of BPs using method (b)), the BP should be {\em localized}, 
i.e., its position between $z(s_0)$ and $z(s_1)$ should be computed. 
A simple way to do this is bisection, 
based on either (a) or (b) as in the original detection. 

Various other methods for detection and localization of BPs have been 
suggested and implemented. Some of these are only suitable 
for small $n_u$, for instance computing {\em all} 
eigenvalues of $\mathcal{A}$, or using bialternate products \cite[\S10.2.2]{Kuznetsov2010}. Other methods 
use so called minimally extended systems, in particular, for 
localization, where typically the dimension of the problem 
is doubled or tripled. Again see \cite[Chapter 10]{Kuznetsov2010}, or for 
instance \cite[Chapter 3]{mei2000}. 

After localization of a BP we want to compute 'the other' branches 
which bifurcate from the branch we continued so far. This branch 
switching is again a predictor-corrector method, and for steady 
bifurcations the main task is to compute tangents to the bifurcating branches. 
We call a BP a {\em simple bifurcation point} if exactly one simple eigenvalue 
goes through zero, such that exactly two (steady) branches intersect at this point. For simple BPs, branch switching is easy, i.e., the bifurcation 
direction follows from explicit formulas, see e.g., \cite{keller77}. 
Similarly, for (simple, i.e., exactly one bifurcating Hopf branch) Hopf 
bifurcations there are somewhat more lengthy formulas 
to obtain approximations of solutions on the bifurcating branches, see, e.g., 
\cite[p531-536]{Kuznetsov2010}. 
To proceed similarly for multiple BPs, we need to 
find isolated solutions of the pertinent 'algebraic bifurcation equations'. 
This may be a difficult problem if it is not clear a priori that the 
 bifurcating branches are determined at some reasonably low order $k$. See, e.g., \cite[\S6.4]{mei2000} for further discussion of this determinacy problem, 
and a general algorithm. However, to the best of our knowledge 
this has not been implemented in full generality in any package, 
and has only recently been partially implemented in \textsc{pde2path} \cite{pattut}. 
For multiple Hopf bifurcations, 
the situation quickly becomes very complicated, see, e.g., 
\cite[Chapter 10]{Kuznetsov2010}, but for instance \textsc{matcont} 
can deal with a large number of cases (in the low dimensional setting). 

\subsection{Continuous symmetries and phase conditions}
\label{sec:cspc}
If (\ref{eq:kinetic}) has continuous symmetries, for instance a boost 
invariance or translational invariance, such that even for fixed 
primary parameter $\lambda$ there is a nontrivial continuous family $\sigma\mapsto \phi(\cdot,\lambda,\sigma)$ of solutions, then the linearization around $\phi$ is always singular. For instance, for a translational invariant (1d) problem  
$\partial_t \phi=F(\phi)$, 
if $F(\phi(\cdot))=0$, then $F(\phi(\cdot+\sigma))=0$ for all $\sigma\in 
\mathbb{R}$, hence $0=\frac{\mathrm{d}}{\mathrm{d}\sigma}F(\phi(\cdot+\sigma))|_{\sigma=0}=\partial_\phi F(\phi)
\partial_x \phi$ such that $\partial_x \phi$ is in the kernel of $\partial_\phi F(\phi)$. 

Such symmetries are approximately inherited by the discretized problem (\ref{e0}). Thus, $\mathcal{A}$ from (\ref{newton}) is almost singular, and the neutral directions must be removed by appropriate phase conditions, for 
which we use the general notation 
\begin{equation}
Q(u,\lambda)=0\in\mathbb{R}^{n_q}, \label{pcgen}
\end{equation}
where now $\lambda=(\lambda_{{\rm org}},\lambda_{{\rm add}})\in \mathbb{R}^{n_p}=\mathbb{R}^{1+n_q}$, 
where $\lambda_{{\rm add}}\in\mathbb{R}^{n_q}$ stands for the required additional parameters. For PDEs, the two most 
common examples are mass conservation 
\begin{equation}
q_1(u,\lambda)= \frac{1}{|\Omega|}\int_\Omega u \, \mathrm{d}\vec r - m_0 = 0 \,,  \label{eq:mass-cons}
\end{equation}
where $m_0$ is a reference mass, 
and phase conservation (here written for the case of translational invariance 
in $x$) 
\begin{equation}
q_2(u,\lambda) = \int_\Omega u\partial_x u^{*}\, \mathrm{d}\vec r = 0 \,,  \label{eq:phase-cond}
\end{equation}
where $u^*$ is a reference profile, and expressions such as $\int u\,\mathrm{d}\vec r$ and $\partial_xu^*$ are 
understood as the discrete analogs of the respective expressions for $\phi$. 
The associated additional parameters $\lambda_{{\rm mass}}$ for $q_1$ and 
$\lambda_{{\rm phase}}$ for $q_2$ are then introduced into (\ref{e0}) in the 
form $\partial_t u=-G(u,\lambda)+\lambda_{{\rm mass}}$ and $\partial_t u=-G(u,\lambda)+\lambda_{{\rm phase}}\partial_x u$, respectively, where the latter corresponds to a transformation to a comoving frame with speed $\lambda_{{\rm phase}}$. \textsc{auto07p} and \textsc{pde2path} have general interfaces 
to add (\ref{pcgen}) to the original problem (\ref{e0}). In detail, by 
augmenting $u$ with the suitable parameters and 
setting the appropriate lines of $M$ in (\ref{e0}) to zero, 
$Q$ can be appended to $G$ in (\ref{e0}), and this is done internally, 
exploiting the flexibility obtained from $M$. 

Examples for mass conservation, phase conservation and a combination of both are given in sec.~\ref{sec:sh-pfc}, \ref{kssec} and \ref{sec:nonvar-tf}, respectively. The conservation of mass (\ref{eq:mass-cons}) in sec.~\ref{kssec}, for example, is achieved by adding a fictitious flux $\varepsilon$ to the time evolution, i.e., $\partial_t \phi = ... + \varepsilon$. This additional free parameter is then automatically kept at zero during continuation with side condition (\ref{eq:mass-cons}).

\subsection{Hopf bifurcation and time periodic orbits}
\def\hoxi{\xi_{\text{H}}}
To compute and continue Hopf orbits, i.e., time-periodic orbits 
with periods near $2\pi/\omega_0$ close to the bifurcation, 
a standard method is to rescale time $t \to t/T$ with an 
unknown parameter $T$, $T=2\pi/\omega_0$ at bifurcation, and consider 
\begin{equation}
M\frac{\mathrm{d}}{\mathrm{d}t} u=-TG(u,\lambda), \quad u(0)=u(1)\,.\label{ho1}
\end{equation}
Since (\ref{ho1}) is autonomous, and hence any translation of a solution 
$t\mapsto u(t)$ is again a solution, we need a phase condition, for 
instance 
\begin{equation}
q:=\int_0^1\left\langle u(t),\dot u_0(t)\right\rangle\,\mathrm{d} t, \label{pca}
\end{equation}
and the step length $p$ from (\ref{esys}) changes to, e.g., 
\begin{equation}
p:=\xi_\mathrm{H}\int_0^1\left\langle u(t){-}u_0(t),u_0'(t)\right\rangle\,\mathrm{d} t+
(1{-}\xi_\mathrm{H})\bigl[w_T(T{-}T_0)T_0'+(1{-}w_T)(\lambda{-}\lambda_0)\lambda_0'\bigr],\label{ala}
\end{equation}
where again $u_0,T_0,\lambda_0$ are from the previous step, $\xi_\mathrm{H}$ and $w_T$ denote 
weights for the $u$ and $T$ components of the unknown solution, 
and the integrals in (\ref{ala}) (and in (\ref{pca})) are discretized based on a time discretization $t_0=0<t_1<\ldots<t_m=1$. For a piecewise 
linear discretization (in $t$), the full unknown 
discrete solution $u$ then is a vector in $\mathbb{R}^{mn_u}$. 
Setting $z=(u,T,\lambda)$ and writing (\ref{ho1}) as $\mathcal{G}(z)=0$ 
we thus obtain the extended system 
\begin{equation}
H(U):=\left(\begin{array}{c} \mathcal{G}(z)\\q(u)\\p(z)\end{array}\right)\stackrel{!}{=}\left(\begin{array}{c} 0\\0\\0\end{array}\right)
\in\mathbb{R}^{mn_u+2}\,.\label{fsa} 
\end{equation}

The (in)stability of and possible bifurcations from a periodic orbit $u$ 
are analyzed via its Floquet multipliers $\gamma$, which are obtained 
from finding nontrivial solutions $(v,\gamma)$ of the variational boundary value 
problem 
\begin{equation}
M\frac{\mathrm{d}}{\mathrm{d}t} v(t)=-T\partial_u G(u(t))v(t),\label{fl1}\quad 
v(1)=\gamma v(0)\,.
\end{equation}
By translational invariance of (\ref{ho1}), there is always the trivial multiplier $\gamma_1=1$, 
and $u_H$ is (orbitally) stable if all other multipliers have modulus 
less than 1. Moreover, a necessary condition for the bifurcation 
from a branch $\lambda\mapsto u_H(\cdot,\lambda)$ of periodic orbits 
is that at some $(u_H(\cdot,\lambda_0),\lambda_0)$, 
additional to the trivial multiplier 
$\gamma_1=1$ there is a 
second multiplier $\gamma_2$ (or a complex conjugate pair $\gamma_{2,3}$) 
with $|\gamma_2|=1$, see, e.g., \cite[Chapter 7]{seydel} or \cite{Kuznetsov2010}. 
For certain types of $t$ discretizations, the Floquet multipliers can 
be obtained from the Jacobian $\partial_u\mathcal{G}(u_H)$ \cite{Lust00}, see also 
\cite[\S 2.4]{hotheo}, but for large scale problems this becomes expensive, and in general the computation of Floquet multipliers is a difficult problem. 

If the original PDE has continuous symmetries and thus the computation 
of steady (or traveling wave) solutions requires $n_q$ phase conditions 
(\ref{pcgen}), then the computation of Hopf orbits requires suitable 
modifications of these phase conditions. This is in general not straightforward 
since (\ref{pcgen}) is not of the form $\partial_t u=Q(u,\lambda)$ and 
thus cannot simply be appended to (\ref{e0}), if the time steppers 
or boundary value problem (in $t$) solvers cannot deal with singular $M$ 
in (\ref{e0}), or (\ref{ho1}). In \S\ref{kssec} we proceed by 
example and consider the Kuramoto-Sivashinsky equation with 
periodic boundary conditions. For the computation of steady 
branches and branches of traveling waves, this requires two phase conditions 
of type (\ref{eq:mass-cons}) and (\ref{eq:phase-cond}), 
and in \S\ref{kssec} we explain how to modify these 
for the computation of Hopf orbits, i.e., for standing waves and 
modulated traveling waves. 

\subsection{Some packages}
The above ideas are the basis for a number of numerical 
continuation and bifurcation packages, with different foci and emphasis. 
Two highly developed and established packages are \textsc{auto07p} \cite{auto} and \textsc{matcont} \cite{matcont}. Both are originally aimed at genuine ODEs, i.e., low-dimensional algebraic problems for the case of steady solutions, and are very powerful 
for this setting. Both have also been applied 
to PDE problems after spatial discretization, 
but this becomes problematic for large $n_u$, mostly because the numerical linear algebra is aimed at ODEs. 

The rather recent package \textsc{pde2path} \cite{p2p14, p2phome} is specifically aimed at PDEs. In particular, it provides 
convenient user interfaces to obtain the form (\ref{e0}) for a rather 
large class of PDE problems in 1d, 2d, and 3d. 
Mesh and FEM space generation, including adaptive mesh refinement, 
work rather automatically, and the application of \textsc{pde2path} to model 
problems of various types with various boundary conditions is explained in 
a number of demo directories and tutorials, available at \cite{p2phome}.
On the other hand, \textsc{pde2path} is as yet restricted to steady bifurcations, 
Hopf bifurcations, periodic orbits and their Floquet multipliers (up to 
medium size problems) and a few codimension-2 problems, but does not yet deal with, e.g., 
secondary bifurcations from Hopf orbits, and thus does not yet provide the same 
completeness and generality for PDEs that \textsc{auto07p} and \textsc{matcont} provide for 
genuine ODE problems. 

Other software used for large scale problems include
\textsc{cl\_matcont} \cite{clmatcont}, which has a focus on invariant
subspace continuation that makes it suitable for larger scale
computations \cite {bindel14}, or \textsc{coco} \cite{cocobook} which
is a general toolbox and for instance in \cite{cocofem} has been
coupled with \textsc{Comsol} for a PDE problem. See also \textsc{loca}
\cite{loca} or \textsc{oomphlib} \cite{oomph} for continuation and
bifurcation tools (libraries) aimed at PDEs.  Concerning the
continuation of periodic orbits, the collocation method used in
\textsc{pde2path} may not be efficient for large scale problems, i.e.,
for more than 30.000 DoF in space, combined with more than 20 DoF in
time. For such problems, shooting methods appear to be more
appropriate, see, e.g., \cite{SanGM2013} for impressive results.  On
the other hand, these and many other continuation/bifurcation results
for PDEs in the literature, see also \cite{Detal14, NSan2015} for
reviews, seem to be based on custom made codes, which sometimes do not
seem easy to access and to modify for nonexpert users, although
\cite{SN16} provides an excellent review of recipes for continuation
based on time steppers and shooting methods.

In this contribution we aim to portray what can be done 
for problems of type (\ref{eq:kinetic}) with \textsc{auto07p} 
and \textsc{pde2path}. Implementation details for the 
\textsc{pde2path} examples can be found in the appendix of \cite{EngelnkemperPHD} and the tutorials 
at \cite{p2phome}, which in particular illustrate that 
all the examples can be treated in a convenient unified way, which 
require only a few user-provided \textsc{matlab} functions. 


\section{Steady states of Allen-Cahn- and Cahn-Hilliard-type equations} \label{sec:ac-ch}

\subsection{Model}

The first and most simple examples we are examining are the Allen-Cahn (AC) equation and the Cahn-Hilliard (CH) equation. The AC equation is obtained by neglecting mass-conserving and nongradient contributions in Eq.~(\ref{eq:kinetic:full}) and by introducing a constant mobility in the nonmass-conserving flux:
\begin{equation}
Q_\mathrm{c} = \vec{j}_\mathrm{c}^\mathrm{ng} = \mu_\mathrm{nc}^\mathrm{ng} = 0 \quad \mathrm{and} \quad Q_\mathrm{nc} = 1\,.
\end{equation}
The CH equation is obtained by neglecting nonmass-conserving and nongradient contributions in Eq.~(\ref{eq:kinetic:full}) and introducing a constant mobility in the mass-conserving flux:
\begin{equation}
\mu_\mathrm{c}^\mathrm{ng} = \vec{j}_\mathrm{c}^\mathrm{ng} = Q_\mathrm{nc} = \mu_\mathrm{nc}^\mathrm{ng} = 0 \quad \mathrm{and} \quad Q_\mathrm{c} = 1\,.
\end{equation}
The energy functional 
\begin{equation}
\mathcal{F}[\phi] = \int_\Omega\frac{\sigma}{2}\left(\nabla \phi\right)^2 + f(\phi) - \mu \phi \,\mathrm{d}\vec r\,. \label{eq:AC_energy}
\end{equation}
is based on Eq.~(\ref{eq:energyAC}) and consists of the sum of an interfacial term $\sim (\nabla \phi)^2$, a bulk energy $f(\phi)$ and an additional term $-\mu \phi$ where $\mu$ corresponds to an external field or chemical potential. The bulk energy can have different forms, here, a simple double well potential
\begin{equation}
f(\phi) =  - \frac{1}{2}\phi^2 + \frac{1}{4}\phi^4
\label{eq:AC_energy_local}
\end{equation}
is employed. An example for a more complicated potential is described is \S\ref{sec:nonvar-tf}. Determining on the one hand the flux $j_\mathrm{nc}^\mathrm{g}$ in the AC case by variation of the functional (\ref{eq:AC_energy}), the governing equation (\ref{eq:kinetic:full}) becomes a second order PDE that is the evolution equation of the nonconserved concentration field $\phi(\vec r,t)$
\begin{equation}
\partial_t \phi = \sigma \Delta \phi + \phi - \phi^3 + \mu\,, \label{eq:AC_evo}
\end{equation}
i.e., the well-known Allen-Cahn equation. Determining on the other hand the flux $\vec{j}_\mathrm{c}^\mathrm{g}$ in the CH case 
with the energy functional (\ref{eq:AC_energy}) leads to the fourth order PDE that is the evolution equation of the conserved concentration field $\phi(\vec r,t)$
\begin{equation}
\partial_t \phi = -\Delta\left[\sigma \Delta \phi + \phi - \phi^3\right] \, \label{eq:CH_evo}
\end{equation}
that is often investigated in the context of phase separation of a binary mixture \cite{Cahn1965jcp,Lang1992,Doi2013} with discussions of, e.g., solutions on 1d and 2d domains 
in Refs.~\cite{Novi1985jsp,NoPe1993prsesa,ThMF2007pf} and Refs.~\cite{MaMW2007ijbc,MMMW2008rmc,BrFT2012pf}, respectively.

To examine the steady states of eqs.~(\ref{eq:AC_evo}) and (\ref{eq:CH_evo}) one sets $\partial_t \phi = 0$ and obtains
\begin{eqnarray}
0 &=& \sigma \Delta \phi + \phi - \phi^3 + \mu \label{eq:AC_steady} \\
\mathrm{and} \quad 0 &=&-\Delta\left[ \sigma \Delta \phi + \phi - \phi^3\right]\,, \label{eq:CH_steady}
\end{eqnarray}
respectively. Further, eq.~(\ref{eq:CH_steady}) can easily be transformed into eq.~(\ref{eq:AC_steady}) by performing two spatial integrations. Thereby the first integration constant is zero for no-flux boundaries
and the second integration constant is denoted $\mu$. Importantly, for eq.~(\ref{eq:AC_steady}) as steady AC equation, $\mu$ is the imposed external field or chemical potential, but for eq.~(\ref{eq:AC_steady}) as steady CH equation, $\mu$ takes the role of a Lagrange multiplier for mass conservation. This implies that $\mu$ is directly (but nonlinearly) related to the mean concentration
\begin{equation}
\phi_0 = \frac{1}{\Omega}\int_\Omega \phi \,\mathrm{d}\vec r \,. \label{eq:AC_meanc}
\end{equation}
The different meaning of $\mu$ for the steady AC and CH cases implies that different types of bifurcation curves can be calculated in the two cases and stabilities of the same solutions may differ in the AC and CH context.
\subsection{Continuation} \label{sec:ac-ch-cont}
As eq.~(\ref{eq:AC_steady}) is a second order semilinear equation the implementation in \textsc{pde2path} is straightforward as described in the tutorials given in the appendix of \cite{EngelnkemperPHD}. Here, we use a square with side lengths $l_x = l_y = 32\pi$ with Neumann boundary conditions. As the latter explicitly break the translational symmetry and the mean concentration is not imposed but adjusts with $\mu$ as external parameter, no integral side conditions are needed and there is only one continuation parameter in each continuation run. 

\begin{figure}[htbp] \center
\includegraphics[width = \textwidth]{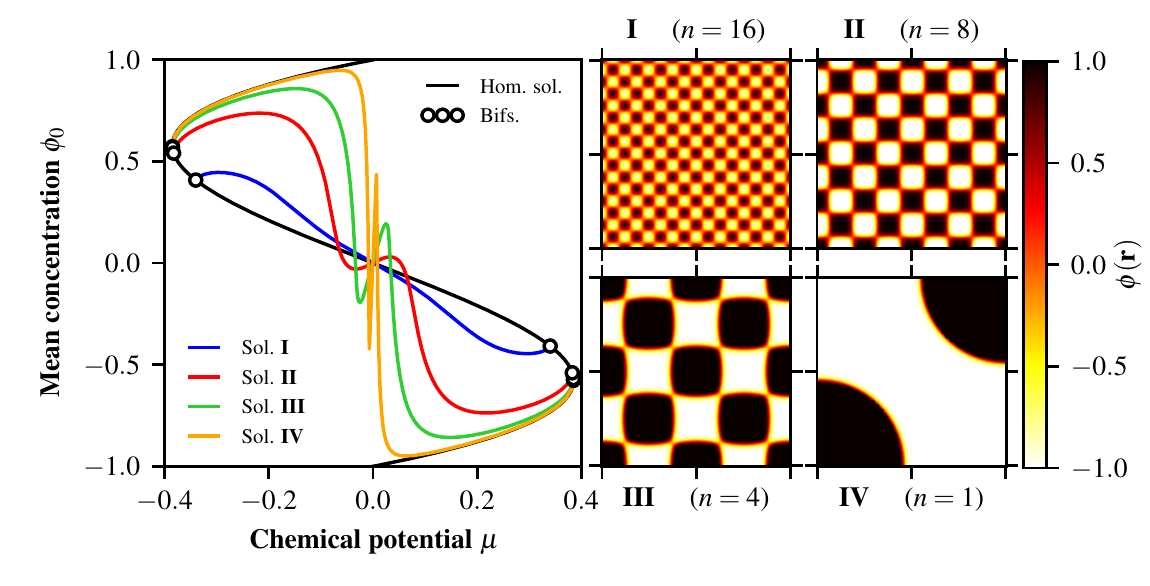}
\caption{(left) Branches of homogeneous and inhomogeneous steady state solutions of the Allen-Cahn equation (\ref{eq:AC_steady}) on a square domain ($l_x = l_y = 32\pi$) with Neumann BC. Shown is the mean concentration $\phi_0$ in dependence of the chemical potential $\mu$. (right) Selected solution profiles for $\sigma = 1.0$ $\mu = 0.1$, $\phi_0 \ne 0$ and different wave numbers $k_x = k_y = n/(16\sqrt{2})$ with $n \in \{1,4,8,16\} $.}
\label{fig:AC_allinone}
\end{figure}
 
We use $\mu$ as continuation parameter, while the mean concentration $\phi_0$ is calculated as solution measure. The obtained bifurcation diagram is shown in fig.~\ref{fig:AC_allinone}.
As starting solution we employ the trivial steady state solution $\phi = \phi_0 = -1$ at $\mu = 0$ and fix the interfacial stiffness to $\sigma = 1$. The first obtained branch (black line in fig.~\ref{fig:AC_allinone}) consists of homogeneous solutions that are analytically known and characterized by
\begin{equation}
0 = \phi_0 - \phi_0^3 + \mu.
\end{equation}
The branch has two saddle-node bifurcations at $(\mu,\phi_0) = \pm (2/3\sqrt{3},-1/\sqrt{3})$. On the sub-branch that connects the saddle-node bifurcations one detects various primary pitchfork bifurcations where inhomogeneous steady states of different wave numbers emerge. The wave numbers $\vec k = (k_x,k_y)^T$ with $k_x = k_y = n/(16\sqrt{2})$ are selected by the employed Neumann BC. These inhomogeneous solutions correspond to phase-separated states where $\phi=\pm 1$ represent the two phases. Note that here we only show as an example a particular pattern type, namely, square patterns with $\vec k$ oriented along the diagonal. One-dimensional patterns, i.e., stripe patterns, and square patterns with wave vectors parallel to the boundaries do also exist on square domains.

We end this section with a remark on the different interpretation of fig.~\ref{fig:AC_allinone} as bifurcation diagram either for the AC or for the CH equation. In the former case,
$\mu$ is the control parameter and $\phi_0$ is the solution measure. Then the bifurcation diagram fig.~\ref{fig:AC_allinone} indicates that along the individual branches of heterogeneous solutions stabilities do not change (i.e., no eigenvalue crosses zero) as there are no saddle-node bifurcations on these branches (and we neglect the possibility of symmetry changing bifurcations). However, in the latter case we have to flip the diagram: now the mean concentration $\phi_0$ is the control parameter and $\mu$ acts as corresponding Lagrange multiplier and in a sense may be considered a solution measure. Looking at fig.~\ref{fig:AC_allinone} in this way (turning the page 90$^\circ$ to the left)
shows that then saddle-nodes occur also on the branches of heterogeneous solutions, i.e., stabilities change along the branches. In other words, the same states may show different stabilities depending on the allowed dynamics - mass-conserving in the CH case and nonmass-conserving in the AC case - as the different dynamics allow for different classes of perturbations.

\subsection{Variational thin-film equation} \label{sec:tf}

Another Cahn-Hilliard-type equation is the thin-film equation of mesoscopic hydrodynamics that describes films and drops of nonvolatile, partially wetting liquid on homogeneous or heterogeneous horizontal substrates \cite{ordb1997rmp,Mitl1993jcis,thie2007,BEIM2009rmp,thie2010jpcm}. The conserved field $\phi(\vec r, t)$ then corresponds to the local film height.%
\footnote{Note that for ultrathin films it can be related to the adsorption per substrate area. This makes it possible to consider transitions between convective dynamics of the bulk of a drop and diffusive dynamics of a molecular adsorption layer (or precursor film) covering the substrate outside the drop \cite{YSTA2017pre}.}
It is obtained from Eq.~(\ref{eq:kinetic:full}) by neglecting nonmass-conserving and nongradient contributions and introducing a cubic mobility in the mass-conserving flux:
\begin{equation}
\mu_\mathrm{c}^\mathrm{ng} = \vec{j}_\mathrm{c}^\mathrm{ng} = Q_\mathrm{nc} = \mu_\mathrm{nc}^\mathrm{ng} = 0 \quad \mathrm{and} \quad Q_\mathrm{c} = \frac{\phi^3}{3}.
\end{equation}
The energy functional is eq.~(\ref{eq:energyAC}) where $\sigma$ now corresponds to the liquid-gas interface tension and
the local energy $f(\phi)$ becomes the wetting potential
\begin{equation}
f(\phi, \vec r) = (1+\xi g(\vec r))\left[-\frac{1}{2}\phi^{-2} + \frac{1}{5}\phi^{-5}\right] \label{eq:TF_bulkenergy}
\end{equation}
which here explicitly depends on the spatial coordinates via a heterogeneity function $g(\vec r)$ encoding $O(1)$ variations and the wettability contrast $\xi$. This models spatial variations of the wettability of the substrate. This particular wetting potential combines two antagonistic power laws \cite{Mitl1993jcis,PiTh2006pf} and allows for the coexistence of a macroscopic drop of equilibrium contact angle $\theta_\mathrm{e}=\sqrt{-2f(\phi_\mathrm{a})/\sigma}=\sqrt{3(1+\xi g)/5\sigma}$ with an adsorption layer of 
height $\phi_\mathrm{a} = 1$ that represents a ``moist'' substrate.\footnote{Consider the discussion of dry and moist case in Ref.~\cite{Genn1985rmp}.} The resulting thin-film evolution equation is
\begin{equation}
\partial_t \phi = -\nabla\cdot\bigg[\frac{\phi^3}{3}\nabla\left(\sigma\Delta \phi + (1+\xi g(\vec r)) \left[-\phi^{-3} + \phi^{-6}\right] \right) \bigg]\,. \label{eq:TF_evo_var}
\end{equation}
In \S\ref{sec:nonvar-tf} a more general form the thin-film equation with nonvariational terms is analyzed. \par To investigate steady states, we set $\partial_t \phi = 0$ and integrate twice to obtain
\begin{equation}
0 = -\sigma \Delta \phi - (1+\xi g(\vec r)) \left[-\phi^{-3} + \phi^{-6}\right] - p \,. \label{eq:TF_steady}
\end{equation}
after nondimensionalization. Here, the integration constant $p$ has the physical dimension of a pressure and in the present mass-conserving CH-type case acts as Lagrange multiplier for mass conservation. Note that in a nonmass-conserving AC-type case it would be an externally imposed vapor pressure.

\begin{figure}[htbp] \center
\includegraphics[width = \textwidth]{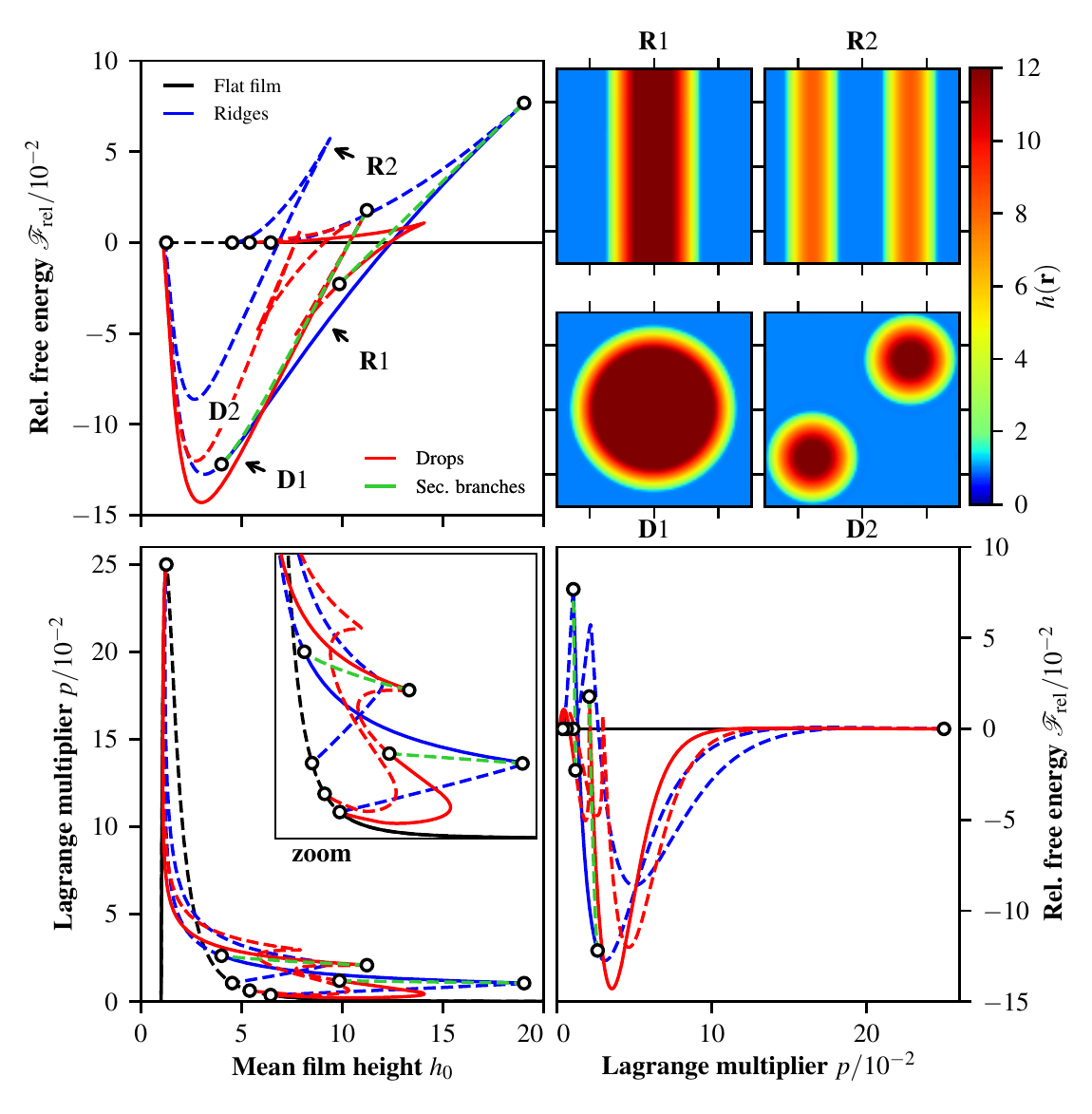}
\caption{Steady state solutions of the thin-film equation for partially wetting liquids on a homogeneous substrate, i.e., solutions of eq.~(\ref{eq:TF_steady}) with $\sigma = 1.0$, $\xi = 0$. The top left panel shows the bifurcation diagram in terms of the energy $\mathcal{F}_\mathrm{rel}$ relative to the energy of a flat film in dependence of the mean film height $\phi_0$ while the bottom left panel gives the corresponding Lagrange multiplier $p$. Linearly stable and unstable solutions (bzgl.~CH-type dynamics) are indicated by solid and dashed lines, respectively. The top right panel shows selected corresponding solution profiles on a domain $\Omega_\mathrm{e} = [-l_x,l_x]\times[-l_y,l_y]$ with periodic boundary conditions with $l_x = l_y = 24\pi$. Finally, the bottom right panel gives the energy $\mathcal{F}_\mathrm{rel}$ in dependence of the pressure $p$ (note that stability is also in this panel indicated for the CH-type dynamics, not the AC-type dynamics).}
\label{fig:TF_allinone}
\end{figure}

With the exception of the explicit spatial dependence of $f(\phi, \vec r)$, the implementation and boundary conditions are equivalent to the ones used in \S\ref{sec:ac-ch-cont}. By extending the numerical domain $\Omega = [-l_x/2,l_x/2]\times[-l_y/2,l_y/2]$ with Neumann boundaries to a physical domain $\Omega_\mathrm{e} = [-l_x,l_x]\times[-l_y,l_y]$ with periodic boundaries  in the case of a homogeneous substrate (i.e., for $\xi = 0$) we determine the bifurcation diagram in fig.~\ref{fig:TF_allinone} using the mean film height $\phi_0$ as control parameter and the energy 
$\mathcal{F}$ relative to the energy of a flat film of identical $\phi_0$ as solution measure 
\begin{equation}
\mathcal{F}_\mathrm{rel}[\phi] = \mathcal{F}[\phi] - \mathcal{F}[\phi_0] = \mathcal{F}[\phi] - \Omega f(\phi_0)\,.
\end{equation}
Note, however, that $p$ acts as Lagrange multiplier that has to be adapted as second continuation parameter because mass is controlled via an integral side condition~(\ref{eq:AC_meanc}). 

The continuation is started with a trivial flat film solution with $\phi = \phi_0 = 1$. Continuing in $\phi_0$ (and adapting $p$) the first run follows the branch of homogeneous films (horizontal line in fig.~\ref{fig:TF_allinone} detecting various pitchfork bifurcations that weakly nonlinear theory identifies as subcritical bifurcations (not shown here, see \cite[chapter 3]{EngelnkemperPHD}).
Again, at the pitchfork bifurcations various nontrivial branches of patterned states of different wave numbers emerge. Here, we focus on branches of solutions with one or two drops (\textbf{D}1 and \textbf{D}2) and with one or two ridges (\textbf{R}1 and \textbf{R}2). As \textbf{D}1 and \textbf{R}1 have the same wave number $k = 1/24$, these patterns bifurcate at the same bifurcations. Furthermore, these drop and ridge branches of solutions of identical wave number are connected by secondary branches emerging from the primary branches at secondary pitchfork bifurcations that break respective symmetries. A linear stability analysis of the flat film shows its instability for $1.26 < \phi_0 < 6.45$ where spinodal dewetting is expected \cite{Thie2003epje}. However, the linearly stable flat film solutions for $6.45 < \phi_0 < 12.43$ coexist with drop and ridge solutions of lower free energy and are therefore metastable. Perturbations of specific forms and amplitude above a certain threshold (related to the threshold solutions on the subcritical part of the respective branches) may still trigger dewetting processes. Depending on the mean film height, the globally stable profiles of minimal free energy are given either by the one-drop (\textbf{D}1) or one-ridge (\textbf{D}1) solution. The transition between them is related to the Plateau-Rayleigh instability of liquid ridges. Figure~\ref{fig:TF_allinone} shows that it is hysteretic and involves metastable regions where finite perturbations are needed to transform ridges into drops or vice versa.
Patterns of larger wave numbers, e.g., the solutions \textbf{D}2 and \textbf{R}2 are always linearly unstable w.r.t. coarsening modes. 

As in section~\ref{sec:ac-ch-cont} we briefly discuss the consequences of the different roles the parameter $p$ takes (i) on the one hand in the context of a mass-conserving CH-type dynamics as the thin-film equation (\ref{eq:TF_evo_var}) and (ii) on the other hand in the context of a nonmass-conserving AC-type dynamics on the same energy functional. In the former case (i), $p$ is the Lagrange multiplier that is adapted during the continuation run. It depends in a nonlinear manner on $\phi_0$ as shown in the lower left panel of fig.~\ref{fig:TF_allinone}. At pitchfork and saddle-node bifurcations in the left panels of fig.~\ref{fig:TF_allinone} the stabilities of the various solutions with respect to mass-conserving perturbations change. However in the latter case (ii) of a nonmass-conserving dynamics (condensation/evaporation), other perturbations are allowed and therefore the stabilities change. In this case, $p$ controls the behavior as external field and $\phi_0$ adapts, i.e., the bottom left panel of fig.~\ref{fig:TF_allinone} can be used as bifurcation diagram if looked at from the left. The bottom right panel gives the corresponding bifurcation diagram showing the energy $\mathcal{F}$ in dependence of the imposed $p$. Again, as in section~\ref{sec:ac-ch-cont}, this change in perspective results in different stability properties of the same solutions, indicated by the absence of saddle-node bifurcations in the bottom right panel of fig.~\ref{fig:TF_allinone}.

\begin{figure}[htbp] \center
\includegraphics[width = \textwidth]{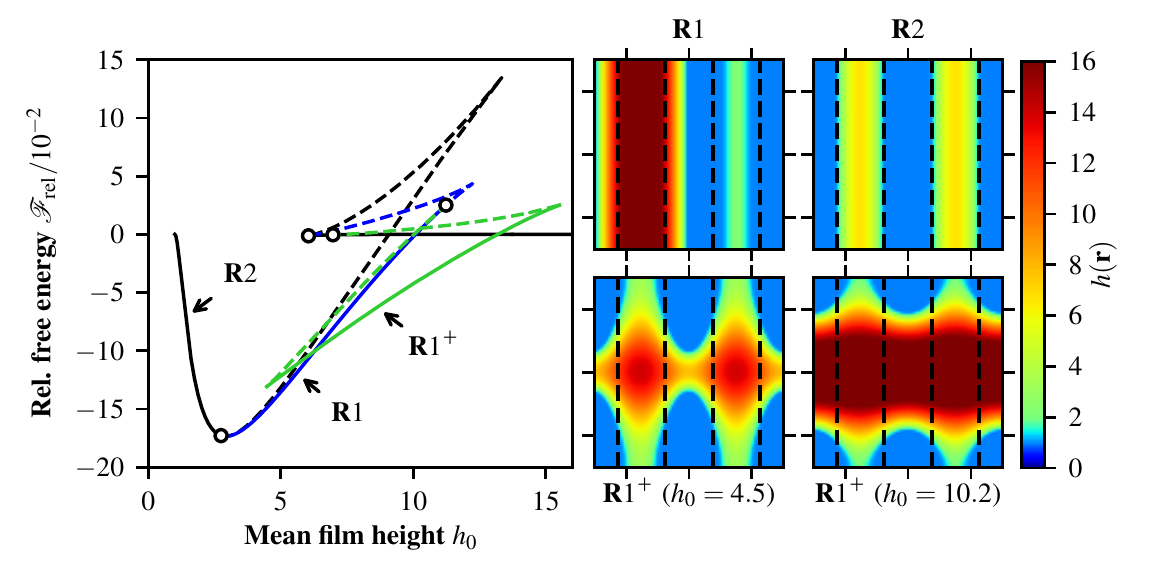}
\caption{Steady state solutions of the thin-film equation for partially wetting liquids on a heterogeneous substrate with two hydrophilic stripes, i.e., solutions of eqs.~(\ref{eq:TF_steady}) and (\ref{eq:TF_het}) with $\sigma = 1.0$, $\xi = 1.0$ and $N=2$. The left panel shows the bifurcation diagram in dependence of the mean film height $\phi_0$ employing again the energy $\mathcal{F}_\mathrm{rel}$ relative to a flat film solution which, however, is no longer a steady state of the given heterogeneous system. Linearly stable and unstable solutions are indicated by solid and dashed lines, respectively. The right panel shows selected corresponding solution profiles for a domain as in fig.~\ref{fig:TF_allinone}.}
\label{fig:TF_het_allinone}
\end{figure}

However, the stability properties of the solutions and the bifurcation structure in the upper left panel of fig.~\ref{fig:TF_allinone} may also change when the wettability of the substrate is modulated \cite{tbbb2003epje} in the case of the mass-conserving dynamics (\ref{eq:TF_evo_var}).
Employing, for instance, a stripelike heterogeneity function
\begin{equation}
g(\vec r) = -\cos\big(\pi Nx/2l_x + \pi(N-1) \big) \label{eq:TF_het}
\end{equation} 
that models a substrate with $N$ hydrophilic stripes, it is possible to stabilize a solution of $N$ ridges against coarsening. In the $N = 2$ case corresponding to the \textbf{R}2 solutions observed on the homogeneous substrate, we use a wettability contrast $\xi = 1.0$ and generate the bifurcation diagram shown in fig.~\ref{fig:TF_het_allinone}.
For the heterogeneous system, the flat film solution does not exist any more but is replaced by a solution of type \textbf{R}2. In the parameter range $1.0 < \phi_0 < 2.74$ this solution is now linearly stable. For larger mean film heights, solution profiles with one ridge parallel (\textbf{R}1) or perpendicular (\textbf{R}$1^\texttt{+}$) to the wettability pattern are linearly stable and/or of minimal energy. The system is analyzed in more detail in \cite{EngelnkemperPHD}.


\section{Steady states of Swift-Hohenberg equation and Phase-Field-Crystal model} \label{sec:sh-pfc}
\subsection{Model}
Another important example of pattern forming models with a scalar
order parameter field is the Swift-Hohenberg (SH) equation
\cite{SwHo1977pra,CrHo1993rmp}. It is obtained from Eq.~(\ref{eq:kinetic:full}) by
neglecting mass-conserving and nongradient contributions, introducing
a constant mobility in the nonmass-conserving flux, i.e.,
\begin{equation}
Q_\mathrm{c} = \vec{j}_\mathrm{c}^\mathrm{ng} = \mu_\mathrm{nc}^\mathrm{ng} = 0 \quad \mathrm{and} \quad Q_\mathrm{nc} = 1\,,
\end{equation}
and employing the energy functional (\ref{eq:en2}) with
$\kappa=\sigma=1$ and
\begin{equation}
f(\phi) =  \frac{1}{2}(1+r)\phi^2 - \frac{\delta}{3}\phi^3 + \frac{1}{4} \phi^4
\end{equation}
This particular SH equation (also other $f$ are common)
reads\footnote{Note that different sign conventions for $r$ are in use.}
\begin{equation}
\partial_t\phi = -(1+\Delta)^2 \phi - r\phi + \delta\phi^2 - \phi^3\,. \label{eq:SH_evo}
\end{equation}
In contrast, the Phase-Field-Crystal (PFC) model \cite{ELWG2012ap},
aka conserved Swift-Hohenberg (cSH) equation \cite{TARG2013pre},  is
obtained by neglecting nonmass-conserving and nongradient
contributions in Eq.~(\ref{eq:kinetic:full}) and introducing a
constant mobility in the mass-conserving flux, i.e.,
\begin{equation}
\mu_\mathrm{c}^\mathrm{ng} = \vec{j}_\mathrm{c}^\mathrm{ng} = Q_\mathrm{nc} = \mu_\mathrm{nc}^\mathrm{ng} = 0 \quad \mathrm{and} \quad Q_\mathrm{c} = 1\,.
\end{equation}
It reads
\begin{equation}
\partial_t\phi = \Delta\left[ (1+\Delta)^2 \phi + r\phi - \delta\phi^2 + \phi^3\right]\,. \label{eq:PFC}
\end{equation}
The corresponding stationary equation is as in \S\ref{sec:ac-ch}
derived by setting $\partial_t\phi = 0$ in eq.(\ref{eq:PFC}) and integrating
twice, obtaining 
\begin{equation}
0 = \Delta^2\phi + 2\Delta\phi + (1 + r)\phi - \delta\phi^2 + \phi^3 - \mu + \varepsilon_1\partial_x \phi + \varepsilon_2\partial_y \phi \,.  \label{eq:SH_steady} 
\end{equation}
where again $\mu$ represents the Lagrange multiplier for mass
conservation. However, eq.~(\ref{eq:SH_steady}) also describes steady
states of the SH equation (\ref{eq:SH_evo}). In this case $\mu$
represents an imposed external field or chemical potential and is in
most works implicitly set to zero. To break the translational
invariances in $x$- and $y$-direction in the case of periodic boundary
conditions we implement two respective phase conditions as described in \S\ref{sec:cspc}.
These imply that two additional continuation parameters are needed. We
chose the fictitious advection speeds $\varepsilon_{1,2}$ (see
comoving frame terms in eq.~(\ref{eq:SH_steady})).

The standard stationary SH equation corresponds to
eq.~(\ref{eq:SH_steady}) with $\delta = \mu = 0$
\cite{CrHo1993rmp}. Employing $r$ as control parameter, its trivial
state $\phi=0$ becomes unstable at a supercritical pitchfork
bifurcation where a branch of stable steady spatially periodic states
emerges. As there is no Maxwell point where linearly stable trivial
and periodic states coexist, the system shows no localized
solutions. However, such a Maxwell point exists when choosing either
$\delta \ne 0$ \cite{BuKn2006pre} or $\mu \ne 0$ \cite{TARG2013pre}
and one can discuss the occurrence of aligned or slanted homoclinic
snaking of localized states.

\subsection{Continuation}
In contrast to the steady AC/CH equation~(\ref{eq:AC_steady}), the
steady SH/PFC equation~(\ref{eq:SH_steady}) is a fourth order
semilinear PDE. To implement this equation in \textsc{pde2path} we
have two options: (i) One may split the fourth order equation into a
system of two second order equations by defining a second scalar field
$u=\Delta \phi$ (see~eqs.~(\ref{eq:SH_split1}) and (\ref{eq:SH_split2}) below). (ii) The fourth order right hand side can be implemented directly by restricting the system to periodic boundary conditions.
Here, we follow the second option that includes the periodic boundary
conditions. To break the resulting translational invariance of the
system we employ a phase condition analogue to
eq.~(\ref{eq:phase-cond}), i.e., adding another continuation parameter
beside the principal one. As solution measure we use the norm
\begin{equation}
||\delta\phi|| := \frac{1}{\Omega} \int_\Omega | \phi - \phi_0| \,\mathrm{d}\vec r\,.
\end{equation}

\begin{figure}[htbp] \center
\includegraphics[width = \textwidth]{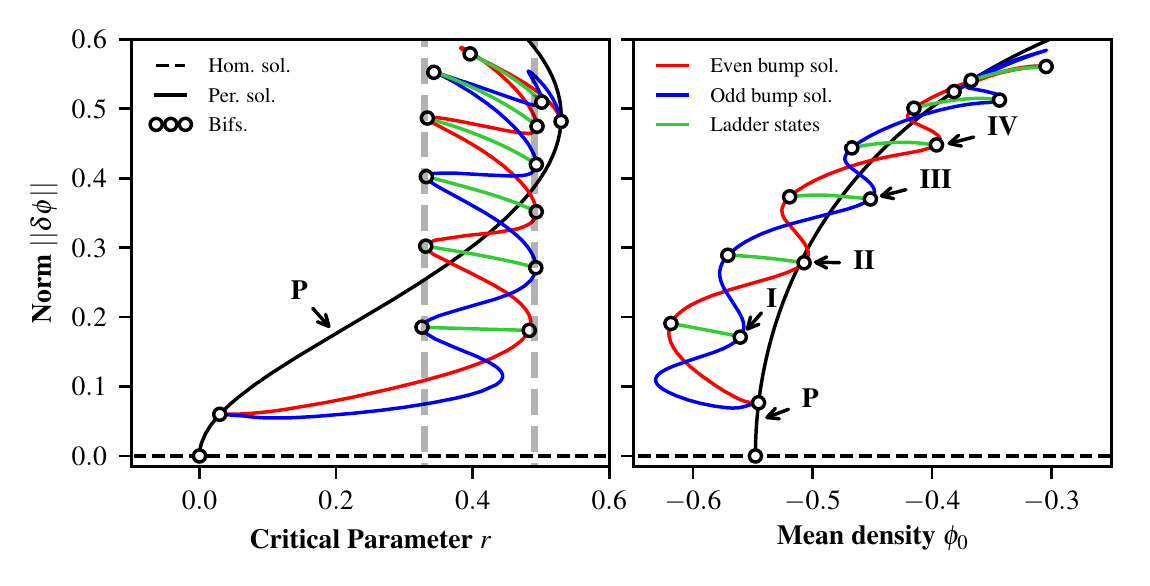}
\caption{(left) Aligned and (right) slanted homoclinic snaking
  branches of localized state of the one-dimensional steady
  Swift-Hohenberg equation (\ref{eq:SH_steady}) for (left) $\delta
  = 2$, $\mu = 0$ and (right) $\delta = 0$, $r = -0.9$, $\mu \ne 0$, respectively.}
\label{fig:SH_branches}
\end{figure}

\begin{figure}[htbp] \center
\includegraphics[width = \textwidth]{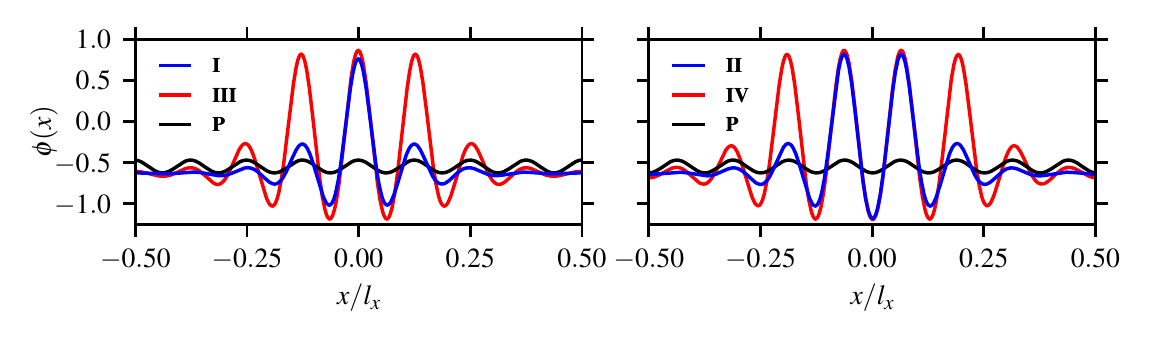}
\caption[]{Profiles of one-dimensional localized states of (\ref{eq:SH_steady}) calculated for $\delta =
  0$, $r = -0.9$ and $\mu \ne 0$ corresponding to the branches shown in
  fig.~\ref{fig:SH_branches}(right). The left panel shows states with
  an odd number of bumps whereas the right panel shows solutions of an
  even number of bumps.}
\label{fig:SH_solutions}
\end{figure}

As starting point we again use a trivial homogeneous solution and
obtain bifurcation diagrams as in fig.~\ref{fig:SH_branches}
employing, e.g., the parameter $r$ or the mean density $\phi_0$ as control parameter. As
first example we look at the (nonmass-conserving) SH equation with
$\delta = 2$ and $\mu = 0$ in 1d, and continue the homogeneous
solution branch in the parameter $r$, detect the primary bifurcations
where the branch of periodic solutions (\textbf{P}) emerges. Following the branch of periodic
states we detect secondary bifurcations where two branches of
localized states bifurcate, namely, of reflection-symmetric states
with even and odd number of ``bumps'', respectively. Typical
solution profiles are shown in
fig.~\ref{fig:SH_solutions}. The second continuation parameter is given by the fictitious advection speed $\varepsilon_1$ which is kept at zero by simultaneously fulfilling the phase condition as given in eq.~(\ref{eq:phase-cond}). The branches of localized states snake
upwards in an intertwined manner, adding a pair of bumps per pair of
saddle-node bifurcations and finally terminate again on the branch
\textbf{P} when the localized structure fills the available space. The
``snake and ladder'' structure is completed by ``runge'' branches of
asymmetric localized solutions that emerge at tertiary bifurcations
and connect the two branches of symmetric localized solutions. For
more details on this see, e.g., refs.~\cite{BuKn2006pre,BuKn2007c}.
Note that the left- and right-hand saddle-node bifurcations are at
respective identical values of the control parameter marked in fig.~\ref{fig:SH_branches}(left) by
vertical dashed gray lines. This
is called aligned snaking.

In contrast, fig.~\ref{fig:SH_branches}(right) shows slanted (or
tilted) snaking where the subsequent respective left and right saddle-node
bifurcations are not vertically aligned. This is, e.g., typical for snaking
branches of localized solutions for systems that involve a
conservation law \cite{BoCR2008pre,Dawe2008sjads,TARG2013pre}. Here, we solve equation
(\ref{eq:SH_steady})  with $\delta = 0$
and use the mean density $\phi_0$ as primary continuation parameter, $\varepsilon$ as further continuation parameter
(translation invariance) and also have to add the Lagrange multiplier $\mu$ as the
third one (as mass conservation gives another integral side
condition). The topology of the bifurcation diagram and sequence of
branches is the same as in fig.~\ref{fig:SH_branches}(left). Solution
profiles are given in fig.~\ref{fig:SH_solutions}.
Note that the linear stability of solutions is not shown here,
cf.~Remark~1 at the end of this section. In general,
the localized states switch from linearly stable to unstable at the
saddle node bifurcations whilst all the ladder states are unstable.
The remark in \S\ref{sec:ac-ch} on the dependence of the stability of
solutions on the character of the considered dynamics also applies here.

\begin{figure}[htbp] \center
\includegraphics[width = \textwidth]{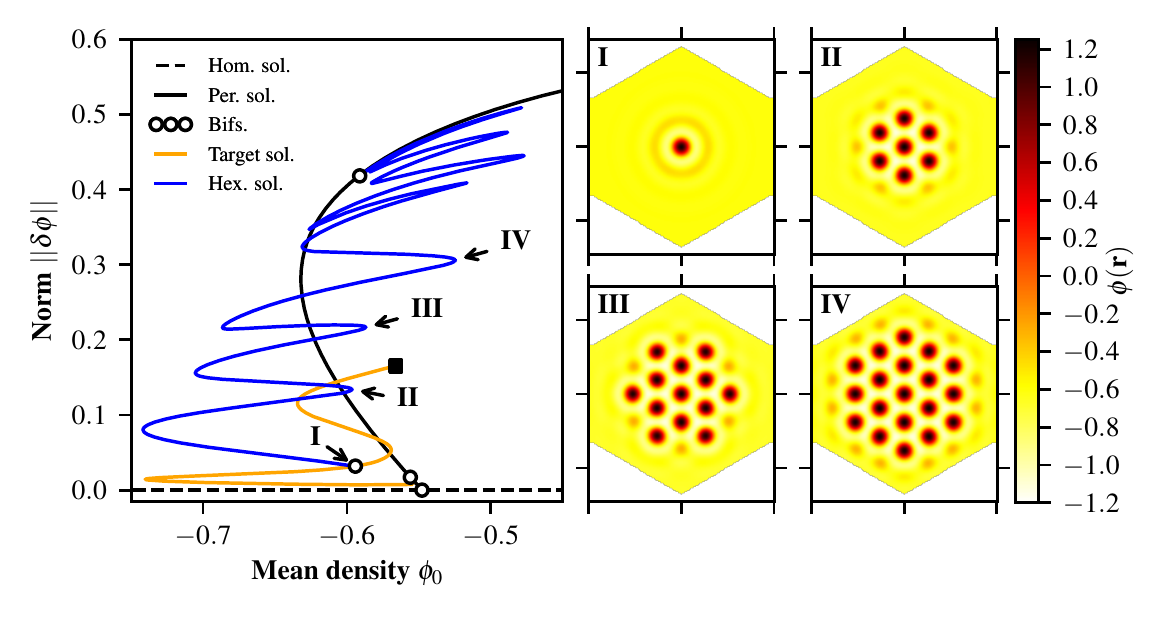}
\caption{ (left) Bifurcation diagram for the steady
  Swift-Hohenberg equation (\ref{eq:SH_steady}) with $\delta = 0$ showing slanted homoclinic
  snaking of localized hexagonal ``crystals'' at the center of a
  two-dimensional hexagonal domain. Shown is the norm as a function of
  the mean density $\phi_0$, while the Lagrange multiplier $\mu$ acts as second continuation parameter. The right panel gives selected profiles at loci indicated by corresponding arrows in the left panel. Again, we set $r = -0.9$.
}
\label{fig:SH_hex_allinone}
\end{figure}

For a two dimensional domain, we can obtain similar solution
branches as shown for a hexagonal
domain in fig.~\ref{fig:SH_hex_allinone} in the mass-conserving case, i.e., using mean
density $\phi_0$ as principal continuation parameter. To increase computational
efficiency we make use of the spatial symmetries and instead of the
full hexagonal domain only use one of twelve equivalent triangles defined by
the vertices $(x,y)=4\pi(0,0)$, $4\pi(0,4/\sqrt{3})$ and
$4\pi(1,\sqrt{3})$ employing Neumann boundary conditions. Therefore we
use option (i) introduced above and split the stationary equation
(\ref{eq:SH_steady}) into the two second order equations
\begin{eqnarray}
0 &=& \Delta \phi - u \label{eq:SH_split1}\\
\mathrm{and} \quad 0 &=& \Delta u + 2u + (1 + r)\phi - \delta\phi^2 + \phi^3 - \mu. \label{eq:SH_split2}
\end{eqnarray}
The first equation (\ref{eq:SH_split1}) defines a second field $u$ as
the laplacian of $\phi$ and the second equation (\ref{eq:SH_split2})
is the steady SH equation written using both fields.

The continuation in $\phi_0$ (adapting the Lagrange multiplier $\mu$
that acts as second continuation parameter) is started with a trivial
homogeneous solution and is switched to a branch of periodic solutions of hexagonal order after the corresponding primary pitchfork bifurcation is detected. In a secondary bifurcation a branch
bifurcates that consists of nearly rotationally invariant localized targetlike
solutions (in a domain that does not show this symmetry). The targetlike structure grows in
extension along the branch. For simplicity we only show the part of this
branch where the interaction of the localized structure with
the Neumann boundaries is negligible (end of shown part marked by black square symbol in fig.~\ref{fig:SH_hex_allinone}(left)). In a tertiary bifurcation localized
hexagonal ``crystals'' emerge. Similar to the one-dimensional case
shown in figs.~\ref{fig:SH_branches} and \ref{fig:SH_solutions}, these
patches grow layerwise along the branch that shows slanted
snaking. Here, an intermediate state (\textbf{III}) is found between
subsequent fully developed hexagons (\textbf{II} and \textbf{IV}),
i.e., it takes four saddle-node bifurcations to add another
layer. Further side-branches exist that are not shown here.
Note that also for two-dimensional domains vertically aligned snaking is found in the case of the
standard nonmass-conserving SH, i.e., eq.~(\ref{eq:SH_steady}) with
$\mu=0$ and $\delta\neq0$ \cite{LSAC2008sjads}. Extended analyses of such localized patterns for reaction-diffusion systems in two-dimensional domains are found in \cite{uwsnak14,w16}.

\begin{remark}\label{cshrem} To also compute on the fly the linear stability
properties of steady states of (\ref{eq:PFC}) we need a slightly different 
splitting instead of (\ref{eq:SH_split1}), (\ref{eq:SH_split2}). We introduce two auxiliary fields $v=\Delta\phi$ and $w=\Delta v$ 
and consider (\ref{eq:SH_evo}) in the form 
\begin{eqnarray}
0&=&\Delta \phi-v\\
0&=&\Delta v-w\\
\partial_t \phi &=&\Delta w+2\Delta v+(1+r)\Delta\phi. \label{consh2} 
\end{eqnarray}
with Neumann-BC, i.e., $\nabla\cdot(v,w,\phi)^T = 0$. 
The FEM formulation of (\ref{consh2}) 
takes the form 
\begin{equation}
\mathcal{M}\partial_t u=-G(u,\mu), 
\end{equation}
where $u=(u_1,u_2,u_3)$ contains the nodal values of $v,w$, and $\phi$, 
respectively, and the mass matrix $\mathcal{M}$ on the left hand side has 
the form
\begin{equation}
\mathcal{M}=\left(\begin{array}{ccc} 0&0&0\\0&0&0\\0&0&M\end{array}\right),
\end{equation}
where $M$ is the one-component 
scalar mass matrix. The crucial point is that the eigenvalue 
problem $\rho \mathcal{M} v=-\partial_u G(u,\lambda)v$ for the linearization around 
some steady solution $u$ then yields the correct discrete eigenvalues 
$\rho$.  
\end{remark}


\section{The Kuramoto-Sivashinsky equation}\label{kssec}
The Kuramoto-Sivashinsky (KS) equation \cite{kura-tsu76,siva77} is a
canonical and much studied nonlinear model for 
long-wave instabilities in dissipative systems, for instance in laminar flame 
propagation, or for surface waves on inclined thin liquid films
\cite{Chan1994arfm}, and is often considered as a model for
spatiotemporal chaos and (interfacial) turbulence \cite{JoKT1990pd,Klia2000jfm,MR815950,MR796268,MR2595887}. Here we 
consider the KS equation in the form 
\begin{equation}
\partial_t \phi=-\sigma\partial_x^4 \phi-\partial_x^2 \phi-\frac 1 2 \partial_x(\phi^2), \label{KS}
\end{equation}
with parameter $\sigma>0$, on the one-dimensional domain $x\in(-2,2)$
with periodic BC. It is obtained from Eq.~(\ref{eq:kinetic:full}) by
neglecting nonmass-conserving contributions and nongradient
contributions to the chemical potential, introducing
a constant mobility in the mass-conserving flux, i.e.,
\begin{equation}
Q_\mathrm{nc} = \mu_\mathrm{nc}^\mathrm{ng} = \mu_\mathrm{c}^\mathrm{ng} = 0 \quad \mathrm{and} \quad Q_\mathrm{c} = 1\,,
\end{equation}
employing the energy functional (\ref{eq:energyAC}) with $f(\phi) =  -\frac{1}{2}\phi^2$,
and the nongradient flux term of type (ii) with $n=2$
(\S\ref{sec:intro}), i.e., $\vec{j}_\mathrm{c}^\mathrm{ng}=-\frac{1}{2}\partial_x(\phi^2, 0)^T$.

\begin{figure}[htbp] \center
\includegraphics[width = \textwidth]{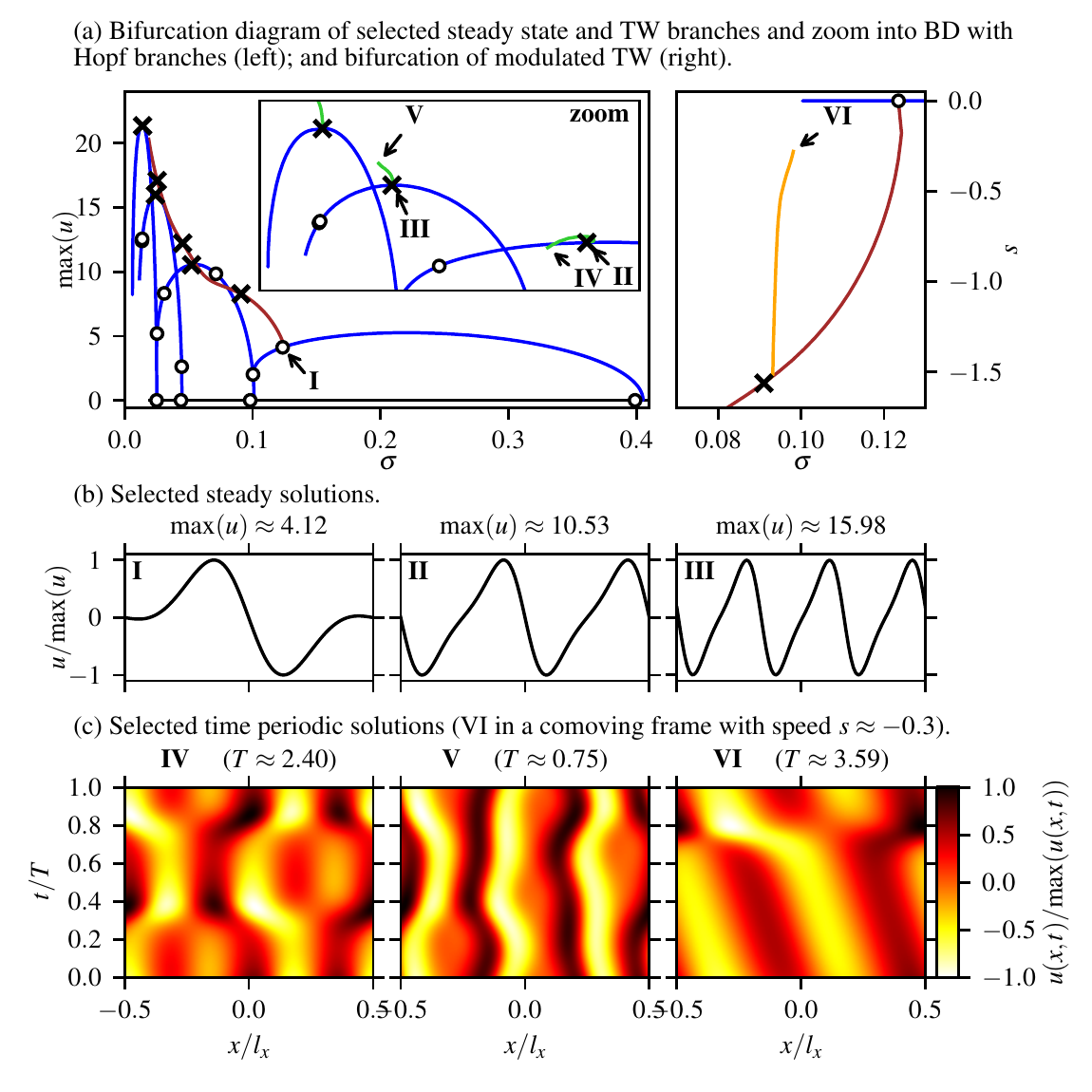}
   \caption{{\small Partial bifurcation diagrams and selected solutions 
for (\ref{KS}). Steady bifurcations (speed $s=0$) and bifurcations to TW ($s\ne 0$) 
indicated by $\circ$, Hopf bifurcations by $\times$.
See main text for detailed comments. \label{ksf1}}
  }
\end{figure}

Equation~(\ref{KS}) with periodic BC is translationally invariant, it has the boost invariance 
$\phi(x,t)\mapsto \phi(x-st)+s$, and thus we need the two phase conditions 
(\ref{eq:mass-cons}) and (\ref{eq:phase-cond}). We therefore modify (\ref{KS}) to 
 \begin{equation}
 \partial_t \phi=-\sigma\partial_x^4 \phi-\partial_x^2 \phi-\frac 1 2 \partial_x(\phi^2)+s\partial_x \phi+\varepsilon\,, 
 \end{equation}
where the wave speed $s$ comes from the comoving frame $x=x-st$, 
and the fictitious influx $\varepsilon$ acts as the additional continuation
parameter related to side condition (\ref{eq:mass-cons}). It is zero (numerically $10^{-10}$) for all computations 
presented here. 
Fixing the mass to $m_0 = \int_\Omega \phi \,\mathrm{d}x = 0$ and employing $\sigma$ as primary continuation parameter,
eq.~(\ref{KS}) shows pitchfork bifurcations (pitchforks of revolution) from the trivial homogeneous solution
$\phi\equiv 0$ to stationary spatially periodic solutions at $\displaystyle
\sigma_k=\left(2/k\pi\right)^2$, $k\in\mathbb{N}$, see bifurcation diagram in
fig.~\ref{ksf1}(a).  Such bifurcations have been studied, e.g., in refs.~\cite{KeNS1990sjam,MR2595887}.
Following the emerging branches with decreasing $\sigma$ we obtain 
secondary Hopf bifurcations on some branches of steady patterns, and for $\sigma\rightarrow 0$ 
the dynamics becomes more and more complicated, making (\ref{KS}) a model 
for (interfacial) turbulence.
Ref.~\cite{vV17} employs time-stepper methods following \cite{SN16} to
determine a fairly complete bifurcation diagram for eq.~(\ref{KS}) (with $\sigma$ in the
range $0.025$ to $0.4$) on a domain $\Omega=(0,2)$ with {\em Dirichlet}
BC, i.e.,
$\phi(0,t)=\phi(2,t)=\partial_x^2\phi(0,t)=\partial_x^2\phi(2,t)=0$. In
particular, many  bifurcations have been explained analytically as
being related to hidden symmetries made visible by  antisymmetrically extending
solutions onto the domain $(-2,2)$ with periodic BC
(also cf.~\cite{CGGK1991lnm}). 

Figure \ref{ksf1}(a) presents basic bifurcation diagrams for (\ref{KS}) as determined with \textsc{pde2path} while
figs.~\ref{ksf1}(b) and~\ref{ksf1}(c) present selected corresponding
profiles of steady and time-periodic solutions, respectively.  As
predicted, at $\sigma_k$ we find supercritical pitchfork bifurcations
where branches of steady periodic solutions emerge. The first one starts out stable at $\sigma_1 = 4/\pi^2 \approx 0.405$, and looses 
stability in another supercritical pitchfork bifurcation at about $\sigma=0.13$ (example solution I) to 
a traveling wave branch (brown), which then looses stability in a Hopf bifurcation (right panel in (a), and example solution (VI)). The
2nd, 3rd and 4th steady state branches gain stability at some rather large amplitude, then loose it again in Hopf bifurcations (II and III for 
the 2nd and 3rd branch). The bifurcating 
Hopf branches consist of standing waves, start out stable but 
become unstable via pitchfork bifurcations. These results all fully agree with those in \cite{vV17} (by antisymmetrically extending the solutions from \cite{vV17}). There, they also compute some further (standing-wave) Hopf 
branches bifurcating in the above pitchforks and period doublings 
from the standing-wave Hopf branches. Here, additionally we have 
traveling waves and Hopf bifurcations to modulated traveling waves, 
with VI just one example. While Fig.~\ref{ksf1} essentially corroborates
the results of ref.~\cite{vV17}, here the results are obtained with a few functions (implementing (\ref{KS}) 
and Jacobians) and commands within the \textsc{pde2path} setup, see \cite[\S5.2]{hotutb} for details. 

For the computation of Hopf orbits we modify eqs.~(\ref{eq:mass-cons}) and 
(\ref{eq:phase-cond}) to 
$$
q^H_1(u(\cdot,\cdot)):=\sum_{i=1}^m \left(\int_\Omega u(t_i,x)\,\mathrm{d} x{-}m_0\right)=0,\quad 
q^H_2(u(\cdot,\cdot)):=\sum_{i=1}^{m-1} \left\langle \partial_x u^*, u(t_i)\right\rangle=0, 
$$
where as before $u$ denotes the spatio-temporal discretization of $\phi$, 
and $\partial_x u^*$ is to be understood in the discrete sense. 
The first equation fixes the average (in $t$) mass to $m_0$, where 
it turns out that the mass is also conserved pointwise (in $t$), while 
the second equation fixes the {\em average} wave speed $s$. 


\section{Convective Allen-Cahn equation} \label{sec:nonvar-ac}

\subsection{Model}

Another example of a nonvariational equation is given by a convective Allen-Cahn (cAC) equation obtained from eq.~(\ref{eq:kinetic:full})
neglecting mass-conserving contributions, introducing a constant mobility in the nonmass-conserving flux, i.e.,
\begin{equation}
Q_\mathrm{c} = \vec{j}_\mathrm{c}^\mathrm{ng} = 0 \quad \mathrm{and} \quad Q_\mathrm{nc} = 1\,,
\end{equation}
and a nongradient parity-breaking rate term $\mu_\mathrm{nc}^\mathrm{ng} =v\partial_x\phi$ that corresponds to a convective term of the form of a comoving frame term.
The energy functional is as for the AC equation [eq.~(\ref{eq:AC_energy})] with 
\begin{equation}
f(\phi) =  \frac{a}{2}[1+\xi g(x)]\phi^2 + \frac{1}{4}\phi^4,
\label{eq:cAC_energy_local}
\end{equation}
i.e., allowing for a spatial variation of the destabilizing term.

Again, restricting ourselves to one-dimensional systems, the resulting equation is
\begin{equation}
\partial_t \phi = \sigma \partial_{xx} \phi - a(1+\xi g(x))\phi - \phi^3 - v \partial_x \phi + \mu \,.
\label{eq:cAC}
\end{equation}
Here, the spatial modulation is given by
\begin{equation}
g(x) = - \tanh\left(x + R_\mathrm{s} \right)  + \tanh\left(x - R_\mathrm{s} \right),
\end{equation}
i.e., $g=0$ for $x\lessapprox -R_\mathrm{s}$ or $x\gtrapprox R_\mathrm{s}$ and $g=-2$ for $-R_\mathrm{s} \gtrapprox x \lessapprox R_\mathrm{s}$.  In this setup for $a>0$, pattern formation is possible in a defined region of length $2R_\mathrm{s}$ (here $= l_x/3$).  The model is meant to describe a very simple spatially extended system where a pinning influence (heterogeneity) and lateral driving force (convective term) compete in the vicinity of a phase transition (double well potential) of a nonconserved order parameter field. It may describe, e.g., a magnetizable foil that is dragged over a window kept at a temperature below the Curie temperature and is at higher temperature elsewhere. Then patterns of spontaneous magnetization may occur in the low-temperature region. The field $\mu$ stands for an external magnetic field.

\subsection{Continuation}

\begin{figure}[htbp] \center
\includegraphics[width = \textwidth]{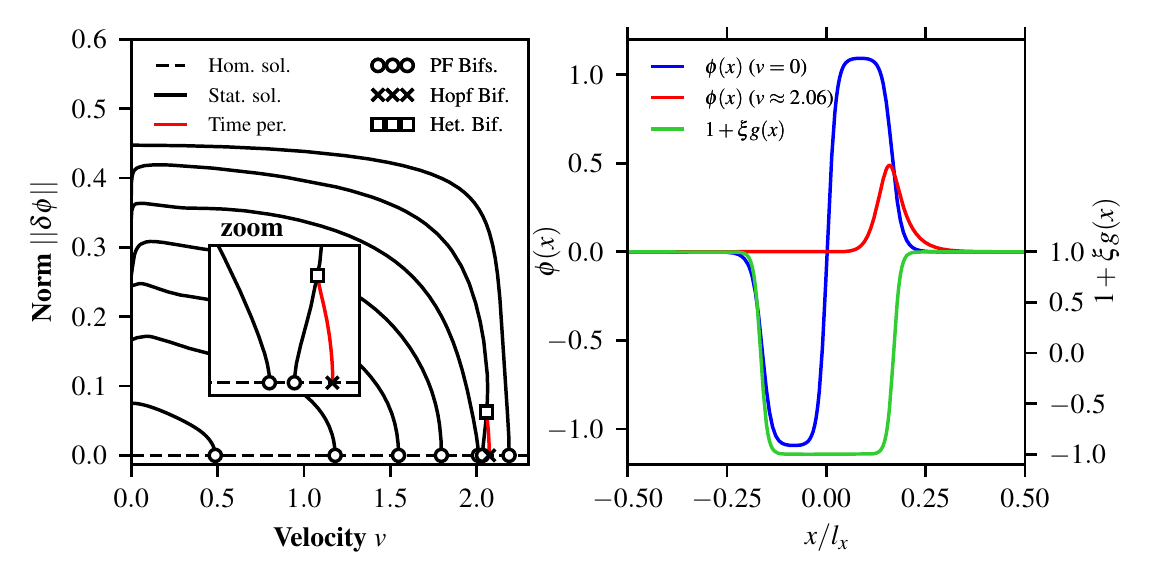}
\caption{(left) Bifurcation diagram and (right) selected steady modulated solution profiles (on second branch from top) for the convective Allen-Cahn equation (\ref{eq:cAC}) for $a = 1.2$, $\mu=0$ and $\sigma = \xi = 1$.
Homogeneous, steady modulated and time-periodic solution branches are shown in the left panel as dashed black, solid black and solid red lines, respectively. Shown is the (time-averaged) norm as a function of the speed $v$. Corresponding space-time plots of time-periodic solutions are given in fig.~\ref{fig:ACAH_tper}.
}
\label{fig:ACAH_stat}
\end{figure}

\begin{figure}[htbp] \center
\includegraphics[width = \textwidth]{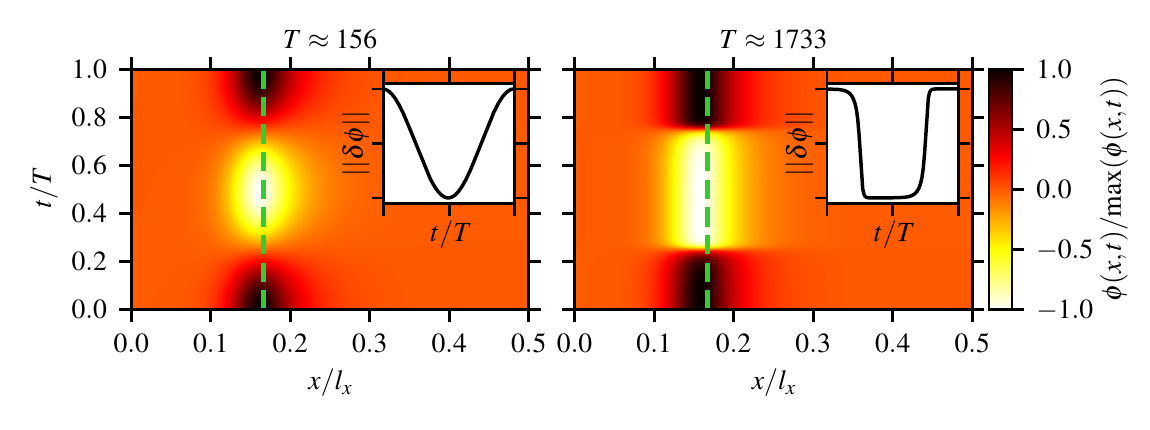}
\caption{Space-time plots of time-periodic states on the branch of time-periodic solutions shown as solid red curve in fig.~\ref{fig:ACAH_stat}(left). The left panel shows a harmonic time dependence close to the Hopf bifurcation while the right panel shows a strongly unharmonic oscillation resembling switching behavior close to the global (heteroclinic) bifurcation.}
\label{fig:ACAH_tper}
\end{figure}

For our one-dimensional problem we use a domain $\Omega = [-l_x/2,l_x/2]$ with $l_x = 60$ and periodic boundary conditions. In contrast to \S\ref{sec:sh-pfc} the explicit spatial dependency induced by the heterogeneity $g(x)$ breaks the translational symmetry, i.e., a phase condition is not needed for continuation. We fix $\sigma = \xi = 1$, $\mu=0$ and use the speed $v$ as the primary (and here only) continuation parameter.

The obtained bifurcation diagram for $a = 1.2$ is given in the left panel of fig.~\ref{fig:ACAH_stat} while the right panel gives selected steady profiles.
The homogeneous solution branch with $\partial_x \phi = 0$ exists for all values of the advection velocity $v$. It is linearly stable at large $v$ and with decreasing $v$ becomes more and more unstable at a number of pitchfork bifurcations. All emerging branches of steady spatially modulated states continue till $v=0$ where then a number of such states exist showing structures with a different number of extrema that are confined to the patterning window. Figure~\ref{fig:ACAH_stat}(right) shows the profile with two extrema (blue curve), i.e., the one connected to the bifurcation at the second largest critical value of $v$, together with the heterogeneity profile (green curve). Whilst at $v = 0$ the solution has the odd symmetry $(x,\phi)\to(-x,-\phi)$, with increasing $v$ the structure gets increasingly suppressed and dragged towards the right border of the patterning window (red curve). Time simulations show that only the steady modulated state of largest norm is linearly stable.

Beside the pitchfork bifurcations the trivial branch can also show Hopf bifurcations. Here, at $a = 1.2$ one Hopf bifurcation is detected at about $v=2.07$ where a branch of time-periodic, spatially modulated solutions emerges. Close to onset they correspond to harmonic oscillations of the stationary solutions shown in fig.~\ref{fig:ACAH_stat} (right panel). An example is given in the space-time plot of fig.~\ref{fig:ACAH_tper}(left). Moving away from onset one finds that the oscillation becomes increasingly unharmonic till finally it resembles abrupt and switches between the two steady modulated solutions that correspond to the nearby branch. \footnote{Note that in the representation with the norm as solution measure each branch of inhomogenous steady states actually represents two branches related by symmetry  $\phi(x) \rightarrow -\phi(x)$.} Decreasing $v$, the overall time period $T$ increases until it diverges when the time-periodic solution ends in a global (heteroclinic) bifurcation on both symmetry-related branches of steady solutions (see inset of Figure~\ref{fig:ACAH_stat}(left)).

Note that changing $a$ one can adjust the number of branches of steady modulated solutions: starting at $a=0$ with increasing $a$ more and more pitchfork bifurcations appear at small $v$ and move towards larger $v$.
Eventually, Hopf bifurcations appear together with heteroclinic bifurcations in codimension-2 Takens-Bogdanov bifurcations at the primary pitchfork bifurcations. Also the pitchfork bifurcations may interact and annihilate, thereby producing branches of steady modulated solutions not connected to the trivial branch (not shown).


\section{Convective Cahn-Hilliard equation} \label{sec:nonvar-ch}

\subsection{Model} \label{sec:nonvar-ch-mod}

There exists a variety of systems described by convective
Cahn-Hilliard (cCH) equations. They are obtained from eq.~(\ref{eq:kinetic:full})
neglecting nonmass-conserving contributions and nongradient chemical potentials, introducing a constant mobility in the mass-conserving flux, i.e.,
\begin{equation}
Q_\mathrm{nc} = \mu_\mathrm{nc}^\mathrm{ng} =\mu_\mathrm{c}^\mathrm{ng} = 0 \quad \mathrm{and} \quad Q_\mathrm{c} = 1\,,
\end{equation}
and a nongradient flux term $\vec{j}_\mathrm{c}^\mathrm{ng}=(v\phi,0)T$ corresponding to a convective term in the form of a comoving frame term.
Employing the energy functional
(\ref{eq:AC_energy}) with (\ref{eq:AC_energy_local}) as for the standard CH equation, one has for a one-dimensional geometry
\begin{equation}
\partial_t \phi = -\partial_{xx} \left[ \sigma \partial_{xx} \phi + \phi - \phi^3- \mu g(x) \right] - v \partial_x \phi
\label{eq:cch1}
\end{equation}
where $g(x)$ is as in \S\ref{sec:nonvar-ac} a spatial heterogeneity
\begin{equation}
g(x) = - \frac{1}{2}\left[1+ \tanh\left(\frac{x - x_\mathrm{s}}{l_\mathrm{s}} \right) \right]\,.
\end{equation}
It again implies that the driving comoving frame term can not be
removed by a coordinate transition as a particular frame of reference
(laboratory frame) is selected by the heterogeneity.

Such equations are relevant for phase separation in dragged-plate
systems like occurring in Langmuir-Blodgett transfer of surfactant
monolayers from the surface of a bath onto a moving plate
\cite{SpCR1994el,CLZH2006jpcb}.  In this context, $\mu g(x)$ is a
space-dependent external field that models the
interaction between the surfactant monolayer and the substrate. It
describes, for instance, the substrate-mediated condensation \cite{SpCR1994el} that
occurs when the surfactant layer comes close to the substrate, i.e.,
in the transition region between the bath and the withdrawing plate with speed $v$ (for details see
Refs.~\cite{KGFT2012njp,KoTh2014n}).  With other words, $g(x)$ is
responsible for a space-dependent tilt of the double-well potential
$f(\phi)$ (see Eq.~(\ref{eq:AC_energy_local}).

There, in certain ranges of parameters like plate velocity and
surfactant concentration, stripe patterns result that can be
perpendicular or parallel to the direction of plate motion. The
stripes are related to the first order phase transition in the
surfactant layer that is triggered by the substrate-mediated
condensation effect \cite{SpCR1994el,CLZH2006jpcb}. Closely related
(coupled) thin-film equations are studied in the context of dip
coating with simple \cite{SZAF2008prl,ZiSE2009epjt,GTLT2014prl}
(Landau-Levich systems) or complex \cite{WTGK2015mmnpb} liquids and,
in general, for solute deposition at receding contact lines
\cite{frat2012sm,DoGu2013el,Thie2014acis}.

Beside the presented convective CH equation (\ref{eq:cch1}) another class of
such equations exist. They consist of Eq.~(\ref{eq:CH_evo}) with an
additional nonlinear driving term
$\sim\partial_x\phi^2\sim\phi\partial_x\phi$, i.e., $j_c^{ng}\sim (\phi^2, 0)^T$ \cite{ZPNG2006sjam} and
model phase separation in driven systems like, for instance,
phase-separating systems with concentration gradients that cause
hydrodynamic motion \cite{GoPi2004c} or the faceting of growing
crystal surfaces \cite{GNDZ2001prl}. In this case the system is
translation invariant.

However, here we only consider the first type of convective CH
equation [eq.~(\ref{eq:cch1})] and emphasize that it may be seen as a
generic model for many systems where a pinning influence, like a
boundary or/and a heterogeneity, competes with a lateral driving force
that keeps the system permanently out of equilibrium (here the plate
motion) in the vicinity of a first order phase transition involving a
conserved quantity (e.g., a phase transition involving a density
change or a wetting transition). As a result one expects the
bifurcation structure of this type of modified Cahn-Hilliard model to
be of interest for a wider class of systems.

\subsection{Continuation} \label{sec:nonvar-ch-cont}

For the numerical continuation
of steady and time-periodic solutions of Eq.~(\ref{eq:cch1}) in a
domain of size $L$ with BC
\begin{equation}
\phi(0) = \phi_0, \quad \partial_{xx}\phi(0) = \partial_x \phi(L) = \partial_{xx}\phi(L)=0 \label{eq:redbc}
\end{equation}
the continuous system is spatially discretized onto an
equidistant grid of $N$ points thereby approximating
the PDE by a dynamical system consisting of $N$ coupled ODEs. 
In Ref.~\cite{KoTh2014n} a second order finite difference scheme is
employed to approximate spatial derivatives.  For this large ODE
system ($N=100\dots400$) the package \textsc{auto07p} can be used to
continue fixed points of the dynamical system that correspond to
steady solution profiles $\phi_0(x)$ of the PDE, to detect local
bifurcations of the fixed points, to continue branches of
time-periodic states, detect their bifurcations and to continue
secondary branches of time-periodic states that emerge at
period-doubling bifurcations. The time-periodic states represent the
deposition of regular line patterns. Before Ref.~\cite{KoTh2014n} such behavior
was only determined via time simulations, i.e., only branches of stable periodic
deposition could be detected \cite{KGFT2012njp}. This is to our
knowledge still the status for all other dip-coating systems.

\begin{figure}[htbp]
\centering
\includegraphics[width=0.8\textwidth]{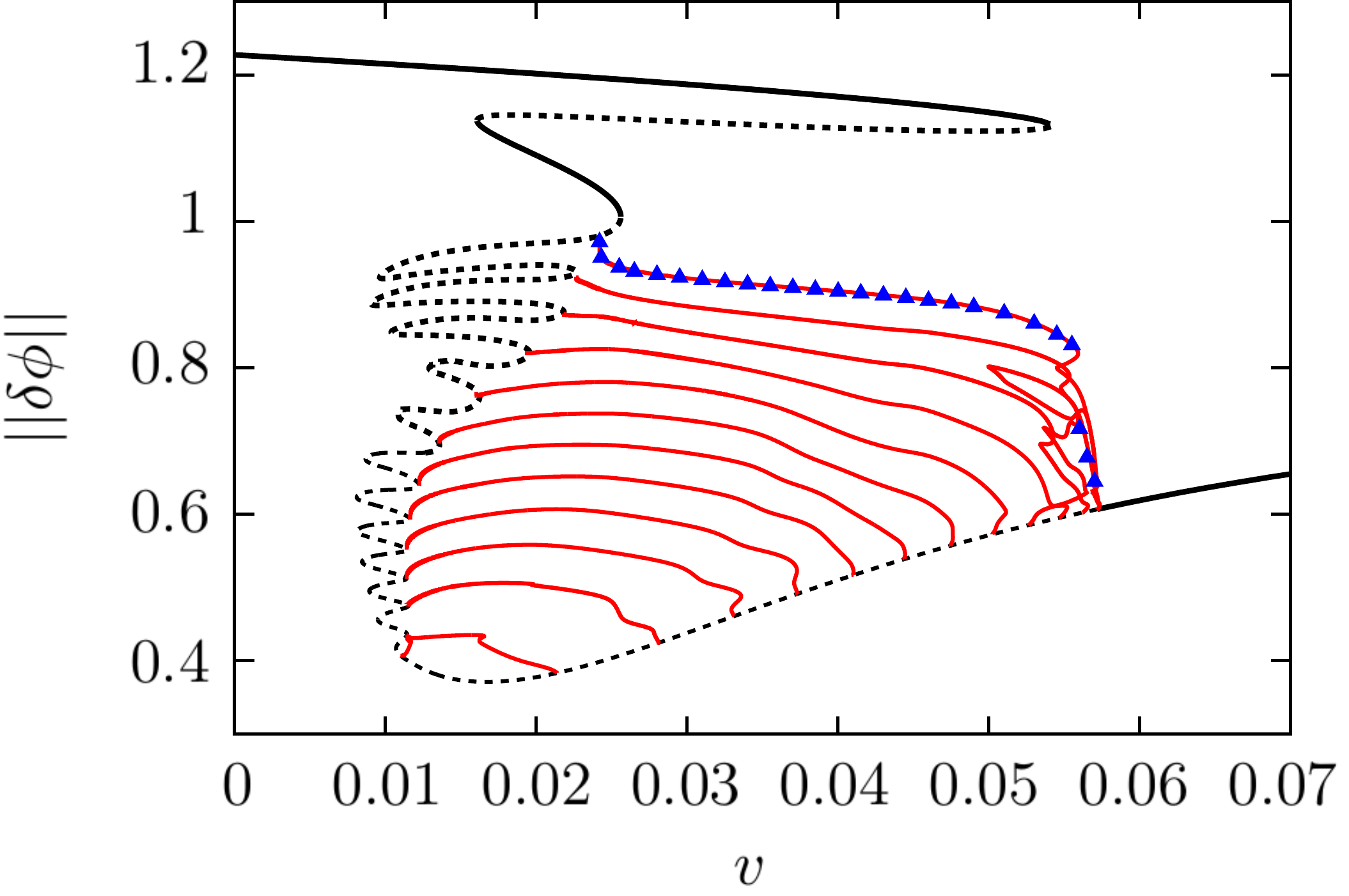}
\caption{Harplike bifurcation diagram for the Langmuir-Blodgett
  transfer system. Shown is the (time-averaged) norm $||\delta \phi||$
  of steady and time-periodic solutions of Eq.~(\ref{eq:cch1}) with
  (\ref{eq:redbc}) in dependence of the dimensionless plate velocity
  $v$ for domain size $L=100$ and $\sigma = 1.0$. For remaining parameters see
  ref.~\cite{KoTh2014n}. The solid and dashed black lines represent
  stable (corresponding to homogeneous transfer) and unstable steady
  profiles, respectively, and the thin solid red lines represent
  time-periodic solutions (corresponding to transfer of stripe
  patterns), all as obtained by numerical path continuation with
  \textsc{auto07p}. The blue triangles correspond to time-periodic
  solutions obtained by direct numerical simulation
  \cite{KGFT2012njp}.  For movies of time-periodic solutions see
  supplementary material of \cite{KoTh2014n}.  Figure adapted from
  Ref.~\cite{KoTh2014n}.  }
  \label{fig:allperiodic}
\end{figure}

However, the various transition scenarios between homogeneous
deposition and various deposition patterns can only be 
understood if the complete structure of the bifurcation diagram is
known including branches of unstable time-periodic states. As an example
we analyze the emergence of stripe
patterns using model (\ref{eq:cch1}) with (\ref{eq:redbc})
employing the described continuation method for 1d substrates. 

In particular, we take the plate velocity as continuation parameter
and obtain the harplike bifurcation diagram shown in
fig.~\ref{fig:allperiodic}. Time-periodic states (corresponding to
transfer of stripe patterns) emerge at low plate velocities through
global (homoclinic or sniper) bifurcations from unstable branches or
saddle-node bifurcations, respectively, of steady profiles that form
part of a snaking family of steady states (for details see caption of Fig.~\ref{fig:allperiodic}).
At high plate velocities the time-periodic solutions emerge through a
number of sub- and supercritical Hopf-bifurcations. Period-doubling
bifurcations are also involved.

 \begin{figure}[htbp]
\centering
\includegraphics[width=0.8\textwidth]{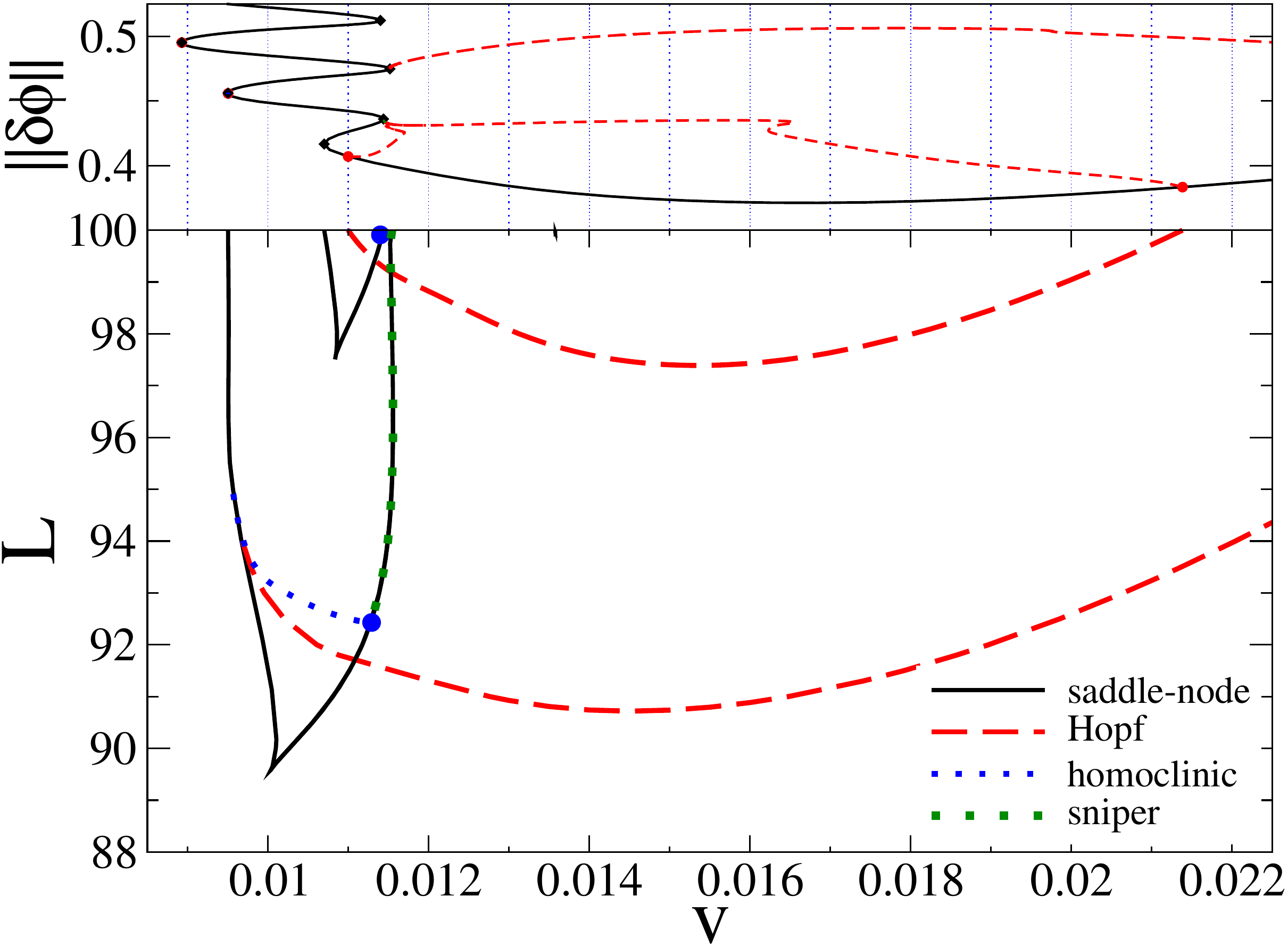}
\caption{Two-parameter continuation of loci of bifurcations: The upper panel shows a zoom onto the tip of the harplike
 bifurcation diagram in fig.~\ref{fig:allperiodic} showing all bifurcations at domain size $L=100$ that
 are continued in the plane spanned by plate
 velocity $v$ and $L$ shown in the lower panel. Note that in the
 chosen parametrization $L$ can be considered as a measure of the
 distance to the critical point of the underlying first order phase
 transition \cite{KoTh2014n}.
In particular, we show loci of four saddle-node bifurcations
  (black solid lines), four Hopf bifurcations (red dashed
  lines), and of the homoclinic [sniper] bifurcation where one of the branches
  of time-periodic states terminates (thin [thick] dotted line). Blue filled circles
  indicate where a homoclinic and a sniper
  bifurcation are created through the collision of a branch of time-periodic states with
  one of steady states. Above the lower filled circle a sniper bifurcation coincides
  with the black solid line of the respective fold. Figure adapted from Ref.~\cite{KoTh2014n}.
}
  \label{fig:allperiodic:onset}
\end{figure}

Beside detecting the bifurcations and following all branches of steady
and time-periodic states, \textsc{auto07p} also allows one to track
the loci of the various bifurcations in appropriate two-parameter
planes and thereby directly determine morphological phase diagrams.
Fig.~\ref{fig:allperiodic:onset} illustrates that this is also possible for the large system of ODEs
resulting from the discretization of the PDE (\ref{eq:cch1}). However,
the continuation sometimes becomes ``fragile'' and at points has to be
restarted with adjusted tolerances etc. With the present methods
it is not yet possible to unambiguously identify the potential
Bogdanov-Takens points where a homoclinic and a Hopf bifurcation
emerge/annihilate (on the leftmost saddle-node bifurcation in
Fig.~\ref{fig:allperiodic:onset}).

For the future, the techniques and investigations should undergo a
major extension towards the case of 2d substrates where transitions
between the various different patterns could not yet been investigated
by continuation methods and are, in general, still rather poorly
understood although many experimental results exist
\cite{Thie2014acis}. Such studies should be performed for reduced
models like Eq.~(\ref{eq:cch1}), but also for the hydrodynamic
thin-film models occurring in the wider class of dip-coating and
receding-contact-line deposition systems
\cite{Thie2014acis, frat2012sm,DoGu2013el,WTGK2015mmnpb}.

For spatially periodic systems, another way of approximating a PDE to
a finite set of ODEs shows better results, i.e., needs fewer
ODEs. Namely, the solution is expressed in the form of a truncated
Fourier series with time-dependent coefficients that are the
independent variables of the resulting dynamical system. Coupling the
package \textsc{auto07p} with a fast Fourier transformation package,
the scheme is used to study depinning droplets on the outside of a
rotating cylinder \cite{LRTT2016pf} and spiral waves in
reaction-diffusion systems \cite{BoEn2007pd}. It may also be applied
to nonlocal (integro-differential) equations, e.g., applying numerical
continuation to obtain bifurcation diagrams for dynamical density
functional theories (DDFT), e.g., for steady states of (potentially
coarsening) clusters of interacting particles in corrugated nanopores
\cite{PoTA2014pre}, and steady and time-periodic states in a model for
the transport of clusters through such pores \cite{PAST2011pre}.


\section{Nonvariational thin-film equation} \label{sec:nonvar-tf}

\subsection{Model}

The previous two sections have analyzed pattern formation in driven AC and CH equations as examples of nonvariational models with nonmass-conserving and mass-conserving dynamics, respectively, using spatially one-dimensional settings. We next investigate a nonvariational thin-film (TF) equation modeling sliding drops on a two-dimensional inclined substrate.
They are obtained from eq.~(\ref{eq:kinetic:full})
neglecting nonmass-conserving contributions and nongradient chemical potentials, introducing a nonconstant nonlinear mobility in the mass-conserving flux, i.e.,
\begin{equation}
Q_\mathrm{nc} = \mu_\mathrm{nc}^\mathrm{ng} =\mu_\mathrm{c}^\mathrm{ng} = 0 \quad \mathrm{and} \quad Q_\mathrm{c} = \frac{\phi^3}{3}\,.
\end{equation}
and a driving nongradient flux term $\vec{j}_\mathrm{c}^\mathrm{ng}=G_0\alpha\frac{1}{3}(\phi^3,0)^T $.
Here, $\phi(\vec r,t)$ is the two-dimensional height profile of a three-dimensional liquid film. To model contributions of Laplace and Derjaguin pressure as well as the hydrostatic pressure proportional to the gravity number $G_0$, we define the energy functional as
\begin{equation}
\mathcal{F}[\phi] =  \int_\Omega\frac{\sigma}{2}\left(\nabla \phi\right)^2 + f(\phi) + \frac{1}{2}G_0\phi^2 \,\mathrm{d}\vec r
\end{equation}
where the nongravitational bulk energy $f(\phi)$ is given by the wetting potential~(\ref{eq:TF_bulkenergy}) in \S\ref{eq:TF_bulkenergy}. Note that by setting $Q_\mathrm{nc} \ne 0$ and $\mu_\mathrm{nc}^\mathrm{g} \ne 0$ a simple evaporation model may be incorporated \cite{thie2010jpcm,Thie2014acis}. The corresponding PDE is then given by 
\begin{equation}
\partial_t \phi = -\nabla\cdot\bigg[\frac{\phi^3}{3}\nabla\underbrace{\left(\sigma \Delta \phi + \left[-\phi^{-3} + \phi^{-6}\right] - G_0 \phi \right)}_{= p} + G_0\alpha\frac{1}{3}(\phi^3,0)^T \bigg]\,. \label{eq:TF_evo}
\end{equation}
where we introduced the pressure $p$.
This equation may also be derived via a long-wave approximation of the Navier-Stokes equations with suitable boundary conditions \cite{ordb1997rmp,thie2007}. As any nonzero inclination angle will cause liquid structures to slide down the substrate, to study stationary moving drops we further transform eq.~(\ref{eq:TF_evo}) into the frame moving with the sliding velocity $U$. The stationary equation is
\begin{equation}
0 = -\nabla\cdot\bigg[\frac{\phi^3}{3}\nabla p + G_0\alpha\frac{1}{3}(\phi^3,0)^T - U(\phi,0)^T \bigg]. \label{eq:TF_comov}
\end{equation}
Defining the new field $u=p +G_0 \phi$, eq.~(\ref{eq:TF_comov}) is split into two second order equations
\begin{eqnarray}
0 &=& u - \sigma \Delta \phi - \left[-\phi^{-3} + \phi^{-6}\right] \\
0 &=& -\nabla\cdot\bigg[\frac{\phi^3}{3}\nabla(u - G_0 \phi) + G_0\alpha\frac{1}{3}(\phi^3,0)^T - U(\phi,0)^T \bigg] + \varepsilon
\label{eq:thifi-twofield}
\end{eqnarray}
Note that for a dragged film system, eq.~(\ref{eq:thifi-twofield}) is still valid in the laboratory frame with $U$ representing driving due to the drag of the moving plate that is inclined by the angle $\alpha$ (analogue to eq.~(\ref{eq:cch1}) in \S\ref{sec:nonvar-ch}).

\subsection{Continuation}

\begin{figure}[htbp] \center
\includegraphics[width = \textwidth]{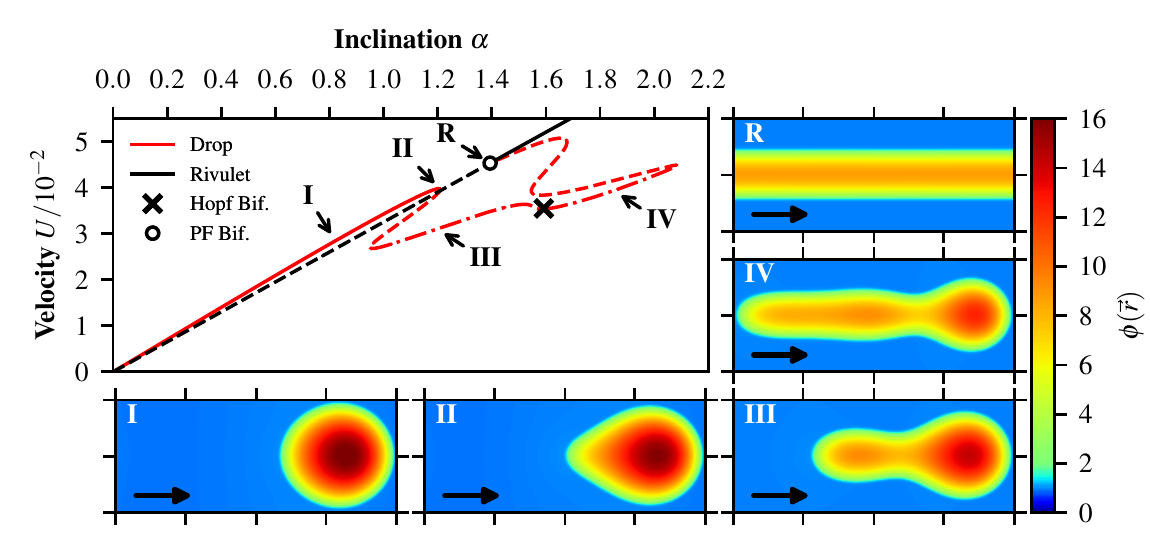}
\caption{(top left) Bifurcation diagram for and (left and bottom) selected corresponding profiles of sliding drops on a domain $l_x\times l_y$ with $l_x = 4l_y = 200$, $\alpha = 0.5$, $\sigma = 1.0$ and Neumann BCs.
The bifurcation diagram gives the drop velocity in dependence of plate inclination. Stable and unstable solutions are given as solid and dashed lines, respectively.}
\label{fig:TF_sliding}
\end{figure}

In contrast to \S\ref{sec:tf}, here, the drop volume or mean film height can not be imposed via a Lagrange multiplier as it only appears in the steady state equation (\ref{eq:TF_steady}) after two integrations, i.e., is not applicable for the present nonvariational system. Here, we use \textsc{pde2path}  to continue the out of equilibrium system (\ref{eq:thifi-twofield}) with integral side conditions for volume conservation (eq.~(\ref{eq:mass-cons})) and the phase condition (eq.~(\ref{eq:phase-cond})) needed due to the periodic boundary conditions in $x$-direction. For the two resulting additional continuation parameters we follow the scheme of \S\ref{kssec}, i.e., we use the sliding velocity $U$ and a fictitious homogeneous influx $\varepsilon$ (that is then kept at zero by the continuation). In $y$-direction we split the physical domain given by $l_x\times l_y = 200\times100$ in half and employ Neumann BC in analogy to \S\ref{sec:tf}.

As starting solution we employ a single steady drop on a horizontal substrate ($\alpha = 0$) similar to profile~\textbf{D}1 in fig.~\ref{fig:TF_allinone}. Employing the inclination angle $\alpha$ as primary continuation parameter we obtain the bifurcation diagram in fig.~\ref{fig:TF_sliding} (top left). Selected corresponding profiles are given in the other panels.
At relatively small $\alpha$, the branch represents linearly stable sliding drops (example profile \textbf{I}) with a nearly linear $U(\alpha)$ dependence. This sub-branch ends in a saddle-node bifurcation at a critical inclination $\alpha_\mathrm{c}$ (example profile \textbf{II}). The branch destabilizes and turns back towards smaller $\alpha$, i.e., a subbranch of linearly unstable profiles coexists with the stable one. At another saddle-node bifurcation the branch turns again towards larger $\alpha$ and passes through some further  saddle-node bifurcations before joining the branch of rivulet solutions of equal volume (example profile \textbf{R}) in a pitchfork bifurcation (in the comoving frame). Note that in the laboratory frame this bifurcation is seen as a traveling wave bifurcation. Along the branch of drop solutions further 
real and/or complex eigenvalues destabilize and stabilize depending of the exact parameter values and details of the model (cf.~\cite{EWGT2016prf,WTEG2017prl}). In all cases, beginning at about the first saddle-node bifurcation, the drop profiles develop an elongation at their rear end (example profiles \textbf{III} and \textbf{IV}).

A detailed analysis of this system is given in ref.~\cite{EWGT2016prf} where the hydrostatic pressure term is neglected. Information from bifurcation diagrams as fig.~\ref{fig:TF_sliding}(top left) may be employed to predict the statistical behavior of ensembles of sliding drops when 'fed into' a Smoluchowski-type statistical model for the distribution of drop volumes in large ensembles of interacting sliding drops \cite{WTEG2017prl}. 

\begin{figure}[htbp] \center
\includegraphics[width = \textwidth]{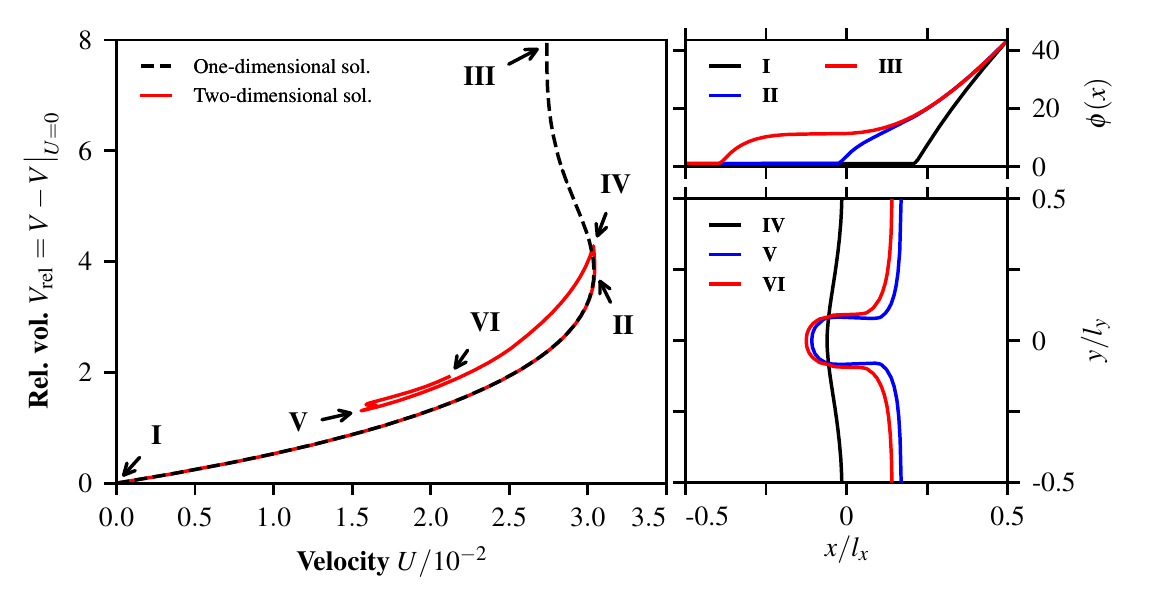}
\caption{Bifurcation diagram (left) and solution profiles (right) for a solid substrate pulled out of a liquid bath with $l_x = 250$, $l_y=220$ and $\sigma = 1.0$. Both, the explicit one dimensional (top right) and the two dimensional case (bottom right) are shown. In the latter case the black, blue and red lines visualize the contact lines of the liquid defined by the condition $\phi(\vec r) = 1.5$.}
\label{fig:TF_men}
\end{figure}

Finally, we briefly discuss the above mentioned related system of dip coating that is also described by eq.~(\ref{eq:TF_comov}): a solid substrate is drawn with velocity $U$ out of a liquid bath deforming the liquid meniscus or dragging a Landau-Levich film onto the moving plate \cite{SZAF2008prl,ZiSE2009epjt,GTLT2014prl,Wilczek2017}. The upper right panel of fig.~\ref{fig:TF_men} shows typical dragged meniscus profiles and acts as a sketch of the system.  To determine steady solutions of the dragged plate problem with eq.~(\ref{eq:thifi-twofield}), the boundary conditions (BC) in $x$-direction have to be changed and both integral side conditions can be dropped as translation symmetry is broken by the BC and the liquid volume is controlled by the influx given through the BC at the side of the bath.  Continuation has been employed in the literature to obtain bifurcation diagrams for one-dimensional substrates using \textsc{auto07p} \cite{GTLT2014prl}, while here we present first \textsc{pde2path} results for fully two-dimensional substrates. Then, transversal instabilities are possible, i.e., the contact line between bath and plate does not need to remain straight.

In our geometry, the bath is on the right hand side, i.e., the plate is withdrawn towards the negative $x$-direction. The particular BC used are on the one hand, Neumann BCs for both fields at the left boundary end ($x = -l_x/2$) where an ultrathin adsorption layer or macroscopic Landau-Levich film is drawn out. On the other hand, at the liquid bath ($x = l_x/2$) we employ Dirichlet BC for the height field $\phi$ (for the shown runs $\phi(l_x/2) \approx 44$) and Neumann BC for the second field $u$, i.e., we impose conditions on $\phi$ and $\partial^3_x \phi$. We employ Neumann BC in the $y$-direction.  Figure~\ref{fig:TF_men}(left) gives the bifurcation diagram for fixed inclination $\alpha = 0.5$ using the plate velocity $U$ as continuation parameter. The dashed line reproduces the result of Ref.~\cite{TsGT2014epje} for a one-dimensional substrate, i.e., the curve for profiles $\phi(x)$ that in a two-dimensional setting represent solutions that are translationally invariant in the transversal $y$-direction.  In contrast, the solid red line gives the fully two-dimensional result. Initially, at small $U$ the two cases are identical, but then slightly above the saddle-node bifurcation a pitchfork bifurcation occurs where solutions with transversally modulated profiles emerge. Examples of ``contact lines'' are shown in the lower right panel of fig.~\ref{fig:TF_men}. At first the modulation is harmonic to then become strongly nonlinear and localized at the center. The branch undergoes further saddle-node bifurcations that shall be further investigated in the future.


\section{Conclusions} \label{sec:summ}

In the present chapter we have given an overview of the application of continuation techniques to a class of nonlinear kinetic equations (\ref{eq:kinetic}) for single scalar order parameter fields.
First, we have introduced a general classification of these equations by distinguishing (i) mass-conserving and nonmass-conserving dynamics, and (ii) gradient (variational) and nongradient (nonvariational) dynamics resulting in the general form eq.~(\ref{eq:kinetic:full}). Furthermore, we have distinguished different types of energy functionals underlying the variational part of the dynamics. As a result the chapter covers Allen-Cahn- and Cahn-Hilliard-type equations (including thin-film equations) as well as Swift-Hohenberg and Phase-Field-Crystal equations -- most of them in their 'pure' variational form and also with various nonvariational extensions that we also classify. Second, we have briefly reviewed the technique of numerical (pseudo-)arclength continuation and the structure of the corresponding numerical algorithms that are employed in the packages \textsc{auto07p} and \textsc{pde2path}. 

In the main part we have given various examples that illustrate how continuation techniques are applied in the investigation of particular physical problems modeled by the introduced equations. For each example we have discussed the specific equation, boundary conditions and integral side conditions and have laid out any resulting implementation issues. Particular emphasis has been put on explaining the number, kind and physical meaning of the various employed continuation parameters. Results have been shown and discussed in the form of selected bifurcation diagrams and solution profiles referring the reader for further details and results to the literature.

In particular, \S\ref{sec:ac-ch} has started with the rather simple example of a pure gradient dynamics of Allen-Cahn type. It has been employed to illustrate the calculation of branches of steady state solutions of PDEs without any side conditions. We have then explained how these branches relate to solutions of  Cahn-Hilliard-type equations that imply a side condition related to mass conservation. Next, we have considered in \S\ref{sec:sh-pfc} branches of steady state solutions of the Swift-Hohenberg and Phase-Field-Crystal equations this time employing periodic boundary conditions. The latter request a phase condition as integral side condition that breaks the translational invariance and implies the usage of an additional continuation parameter that represents a fictitious advection. As the corresponding steady PDEs are of fourth order we have discussed both, the concept of splitting a system in multiple second order equations and directly implementing a fourth order equation in \textsc{pde2path}. Examples have been given in one and two dimensions and the relation between the steady solutions of both equations has been discussed.

The remaining sections have discussed equations with nonvariational parts. First, \S\ref{kssec} has analyzed the Kuramoto-Sivashinsky equation and has introduced the ability of \textsc{pde2path} to detect Hopf bifurcations and to continue emerging branches of time-periodic solutions, namely, travelling and modulated waves. In \S\ref{sec:nonvar-ac} and \ref{sec:nonvar-ch} we have then presented results for nonvariational forms of the Allen-Cahn and Cahn-Hilliard equation, respectively, in cases where an explicit spatial dependency exist. These sections also contain examples for the detection of Hopf-bifurcations and continuation of emerging branches with both, \textsc{pde2path} and \textsc{auto07p}, in part involving period-doubling bifurcations of time-periodic states and global bifurcations. Section \S\ref{sec:nonvar-tf} has then dealt with two nonvariational forms of the thin-film equation describing drops sliding on an incline and a dragged meniscus problem, respectively. 

The chapter has laid out the present state of the art of the application of continuation techniques to the class (\ref{eq:kinetic:full}) of nonlinear kinetic equations for one- and two-dimensional domains. While we have focused on such equations for a single scalar
field, there are many more related equations when allowing for several coupled scalar fields
\cite{FiDi1997pre,Klia1999jfm,crma2009rmp,NaTh2010n,HiKA2012pre,MRTR2012pa,SATB2014c,NSan2015,ThAP2016prf} or indeed coupled scalar, vector and tensor fields \cite{SoVi2001pre,MeLo2013prl,OzHD2016epje}. Complicated problems involving several scalar fields are dealt with in
many of the tutorials and demos coming with \textsc{pde2path}, and research papers using \textsc{pde2path}, in 1d, 2d and 3d, for instance 
\begin{itemize}
\item reaction-diffusion systems in \cite[\S4]{p2p14}, \cite{hotheo} and, mostly ecology related, in  \cite{uwsnak14, SDERMS15, w16, ZUFM17}, including some optimal control aspects \cite{U16}, \cite[\S3.4]{hotheo}; 
\item Bose-Einstein condensates \cite{DS16, DU14}, Rayleigh-Benard convection 
 and related fluid problems \cite[\S5]{p2p14}, \cite{ZHR15}; 
\item Poisson-Nernst-Planck systems, i.e., ionic liquids in \cite{BGUY17, NirG17}; 
\item Laser dynamics systems in \cite{GJ-PRA-17, SJG-PRA-18}.
\end{itemize}
There are a few works where continuation is applied to models that couple scalar fields to vector fields 
\cite{SanGM2013,Detal14} (also consider ref.~\cite{NSan2015,SN16} for more details on the used techniques).

Note that also the continuation of branches 
of time-periodic solutions in 2d and 3d is still somewhat delicate in \textsc{pde2path}, in particular 
with constraints such as mass
conservation and phase conditions. An example is given by the Hopf bifurcation
that has been detected in the sliding drop system shown in fig.~\ref{fig:TF_sliding}. In ref.~\cite{EWGT2016prf} the
dynamics near the bifurcation is already studied by direct time simulations that indicate that the bifurcation is subcritical, but the continuation of the branch quickly 
leads to numerical problems. Here, we may need to switch to shooting based methods as in \cite{SN16}. Also, bifurcations from periodic orbits still await implementation, and the numerical linear algebra can and will be improved at several places.



\section*{Acknowledgements}

DW thanks the Deutsche Forschungsgemeinschaft for support (DFG, Grant
No. Ue60/3-1); HU thanks Jens Rademacher for discussions on suitable
formulations of constraints for Hopf orbits; UT acknowledges funding
by the German-Israeli Foundation for Scientific Research and
Development (GIF, Grant No. I-1361-401.10/2016); and SG acknowledges
partial support by DFG within PAK 943 (Project No. 332704749).  We
thank Daniele Avitabile, Andrew Hazel, Edgar Knobloch, and David Lloyd
for frequent discussions on continuation techniques, bifurcation
theory and pattern formation.


\end{document}